\documentclass[pre,preprint]{revtex4}
\usepackage{graphicx}
\usepackage{dcolumn}
\usepackage{bm}

\begin{document}

\title{ Quantum effects in energy and charge transfer in an artificial photosynthetic complex }

\author{ Pulak Kumar Ghosh$^{1}$\footnote[1]{Author to whom correspondence should be addressed. Electronic mail: pulak@riken.jp},  Anatoly  Yu. Smirnov$^{1,2}$, and Franco Nori$^{1,2}$}

\affiliation{ $^1$ Advanced Science  Institute, RIKEN, Wako, Saitama, 351-0198, Japan \\
$^2$  Physics Department, The University of Michigan, Ann Arbor, MI 48109-1040, USA }

\date{\today}

\begin{abstract}
We investigate the quantum dynamics of energy and charge transfer in a wheel-shaped artificial photosynthetic antenna-reaction center complex.
This complex consists of six light-harvesting chromophores and an electron-acceptor fullerene. To describe quantum effects on a femtosecond time
scale, we derive the set of exact non-Markovian equations for the Heisenberg operators of this photosynthetic complex in contact with a Gaussian
heat bath. With these equations we can analyze the regime of strong system-bath interactions, where reorganization energies are of the order of
the intersite exciton couplings. We show that the energy of the initially-excited antenna chromophores is efficiently funneled to the
porphyrin-fullerene reaction center, where a charge-separated state is set up in a few picoseconds, with a quantum yield of the order of 95\%.
In the single-exciton regime, with one antenna chromophore being initially excited, we observe quantum beatings of energy between two resonant
antenna chromophores with a decoherence time of $\sim$~100 fs. We also analyze the double-exciton regime, when two porphyrin molecules involved
in the reaction center are initially excited. In this regime we obtain pronounced quantum oscillations of the charge on the fullerene molecule
with a decoherence time of about 20 fs (at liquid nitrogen temperatures). These results show a way to directly detect quantum effects in
artificial photosynthetic systems.
\end{abstract}


\maketitle

\section{Introduction}

The multistep energy-transduction process in natural photosystems begins with capturing sunlight photons by light-absorbing antenna chromophores
surrounding a reaction center \cite{Blankenship02,Amerongen00}. The antenna chromophores transfer radiation energy to the reaction center
directly or through a series of accessory chromophores. The reaction center harnesses the excitation energy to create a stable charge-separated
state.

Energy transfer in natural and artificial photosynthetic structures has been an intriguing issue in quantum biophysics due to the conspicuous
presence of long-lived quantum coherence observed with two-dimensional Fourier transform electronic spectroscopy \cite{Engel07,Panitch10}. These
experimental achievements have motivated researchers to investigate the role of quantum coherence in very efficient energy transmission, which
takes place in natural photosystems \cite{Guzik,Plenio,IshizakiPNAS,ChengFleming09,Sarovar}. Quantum coherent effects surviving up to room
temperatures  have also been observed in artificial polymers \cite{Collini09}.  Artificial photosynthetic elements, mimicking natural
photosystems, might serve as building blocks for efficient and powerful sources of energy \cite{Barber09,Larkum10}. Some of these elements have
been created and studied experimentally in Refs.~\cite{gali1,gali2,gust1,gust2,gust3,Imahori}. The theoretical modelling of artificial reaction
centers has been recently performed in Refs.~\cite{GhoshJCP09,SmirnovJPC09}.

\begin{figure}[tp]
\centering
\includegraphics[width=8.6cm]{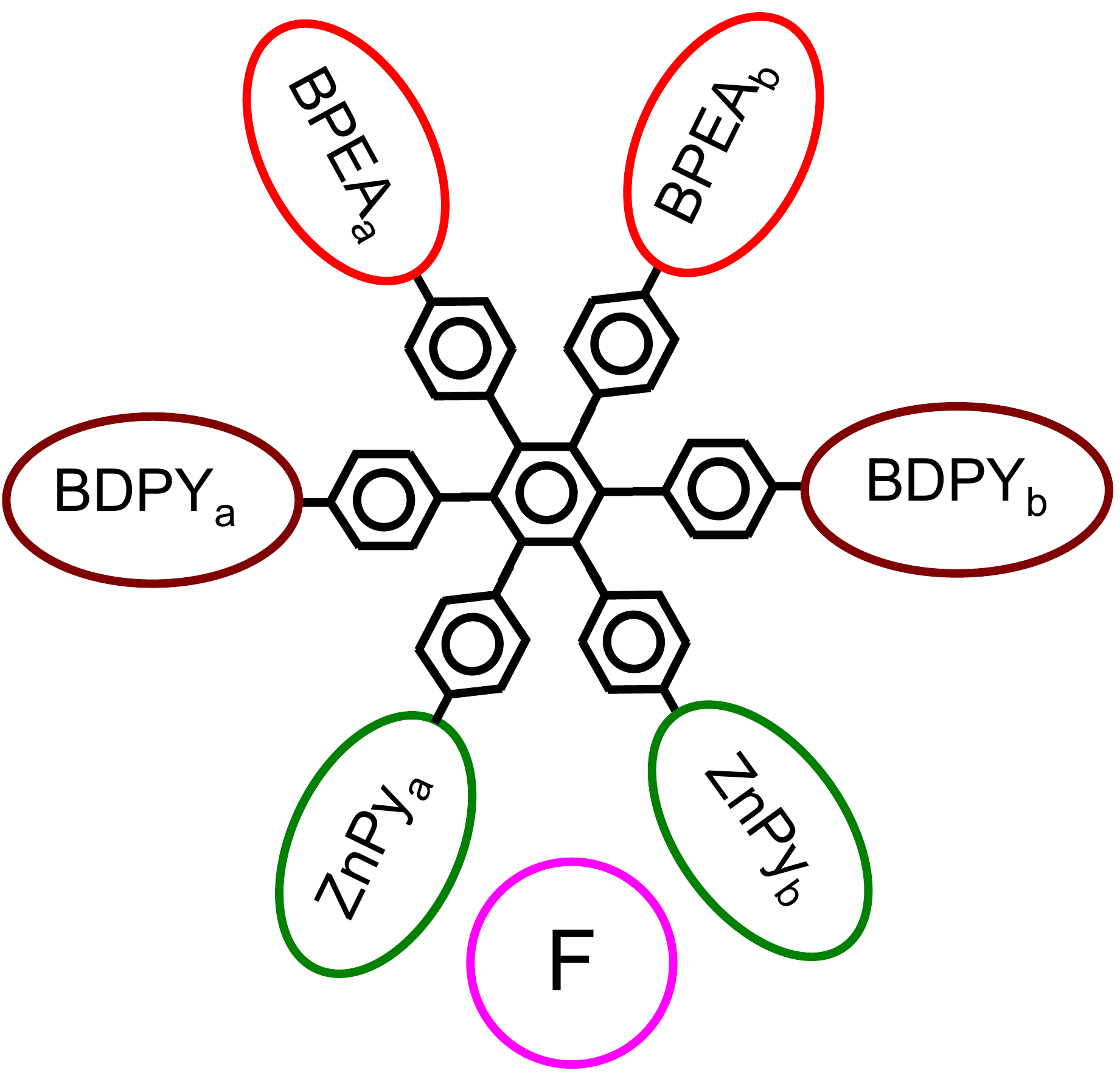}
\caption{(Color online) Schematic diagram of the wheel-shaped artificial antenna-reaction center complex reported in ref \cite{gust3}. We use
the short notation, \emph{BPF Complex}, to denote this photosynthetic device. The antenna-reaction center complex contains six light-harvesting
pigmets: (i) two bis (phenylethynyl)anthracene chromophores, BPEA$_a$ and BPEA$_b$, (ii) two borondipyrromethene chromophores, BDPY$_a$  and
BDPY$_b$,
 and (iii) two zinc tetraarylporphyrin chromophores, ZnPy$_a$ and
ZnPy$_b$. All the chromophores are attached to a rigid hexaphenyl benzene core. In addition to the antenna components, the photosystem contains
a fullerene derivative (F) containing two pyridyl groups, acting as an electron acceptor. The fullerene derivative F is attached to the both
ZnPy chromophores via the coordination of the pyridyl nitrogens with the zinc atoms. For structural details of the BPF Complex we refer
\cite{gust3,gust4}.}
\end{figure}

Here we study energy transfer and charge separation in a wheel-shaped molecular complex (BPF complex, see Fig.~1) mimicking a natural
photosynthetic system. This complex has been synthesized and experimentally investigated in Ref.~\cite{gust3}. It has four antennas - two
bis(phenylethynyl)anthracene (BPEA) molecules and two borondipyrromethene (BDPY) chromophores, as well as two zinc porphyrins (${\rm ZnPy_a}$
and ${\rm ZnPy_b}$). These six light-absorbing chromophores are attached to a central hexaphenylbenzene core. Electrons can tunnel from the zinc
porphyrin molecules to a fullerene F (electron acceptor). Thus, two porphyrins and the fullerene molecule form an artificial reaction center
(${\rm ZnPy_a-F - ZnPy_b}$). The BPEA chromophores strongly absorb around 450 nm (the blue region), while the BDPY moieties have good
absorptions around 513 nm (green region). Porphyrins have absorption peaks at both red and orange wavelengths. Therefore, the BPF complex  can
utilize most of the rainbow of sunlight -- from blue to red photons. It is shown in \cite{gust3} that the absorption of photons results in the
formation of a porphyrin-fullerene charge-separated state with a lifetime of 230 ps; in doing so, excitations from the BPEA and BDPY antenna
chromophores are transferred  to the porphyrins with a subsequent donation of an electron from the excited states of the porphyrins to the
fullerene moiety. This process takes a few picoseconds, suggesting that the excitonic coupling between chromophores is sufficiently strong. The
electronic coupling between the porphyrins and the fullerene controlling tunneling of electrons in the artificial reaction center also should be
quite strong. It should be noted, however, that spectroscopic data ~\cite{gust1,gust2,gust3} show that the absorption spectrum of the BPF
complex  is approximately represented as a superposition of contributions from the individual chromophores with almost no perturbations due to
the links between the chromophores. This means that the chromophores comprising the light-harvesting complex can be considered as individual
interacting units, but not as an extended single chromophore. We can expect that, at these conditions, quantum coherence is able to play an
important role in energy and charge transfer dynamics, manifesting itself in quantum beatings of chromophore populations as well as in quantum
oscillations of the charge accumulated on the fullerene molecule. In principle, these oscillations could be measured  by a sensitive
single-electron transistor, thus providing a direct proof of quantum behavior in the artificial photosynthetic complex. Since these phenomena
occurs at very short time scales (a few femtoseconds), these could be within the reach of femtosecond spectroscopy in the near future. The main
goal of this study is to explore quantum features of the energy and charge transfer in a wheel-shaped antenna-reaction center complex at
subpicosecond timescales.

\section{ Model and Methods}

\subsection{Hamiltonian}

Each chromophore has one ground and one excited state, whereas the electron acceptor fullerene F has just one energy level with energy $E_F$.
 We introduce creation (annihilation) operators, $a_{k}^{\dagger}$ ($a_k$), of an
electron on the $k$th site. The electron population operators are defined as $n_k = a_k^{\dagger} a_k$. We assume that each electron state can
be occupied by a single electron, as spin degrees of freedom are neglected. The basic Hamiltonian of the system has the form:
\begin{eqnarray}
H_0 &=&  \sum _{k} (E_k n_k  + E_{k^{*}} n_{k^{*}}) + E_F n_F + H_C + \sum _{k \neq l} V_{kl} a^{\dagger}_{k^*}a_{k}\;a^{\dagger}_{l}a_{l^*} -
\sum _{\sigma \sigma'} \Delta_{\sigma \sigma'} a^{\dagger}_{\sigma}a_{\sigma'}, \label{H0}
\end{eqnarray}
where the first part incorporates the energies of the electron states (hereafter $k,l$ = BPEA$_a$, BPEA$_b$, BDPY$_a$, BDPY$_b$, ZnPy$_a$,
ZnPy$_b$), and the second term is related to a fullerene energy level $E_F$ with a population operator $n_F = a_F^\dagger a_F$.  The pair
($k,k^{*}$) denotes a ground ($k$) and an excited ($k^{*}$) state of an electron located on the site $k$ with the corresponding energy $E_k
\;(E_{k^{*}})$. The term $H_C$ represents the contribution of Coulomb interactions between electron-binding sites. This term is given in
Appendix A. The fourth term of Eq.~(\ref{H0}) describes excitonic couplings between the chromophores $k$ and $l$. The matrix element $V_{kl}$ is
a measure of an interchromophoric coupling strength. The last term in Eq.~(\ref{H0}) describes the electron tunneling from excited states of the
porphyrin molecules ZnPy$_a$, ZnPy$_b$ to the electron acceptor F characterized by the tunneling amplitudes $\Delta_{\sigma \sigma'}$, where
$\sigma,\;\sigma'$ = ZnPy$_a^{*}$, ZnPy$_b^{*}$, F.

The interaction of the system with the environment (heat bath),
 represented here by a sum of independent oscillators with
Hamiltonian
\begin{eqnarray}
H_{\rm{env}}= \sum _{j}\left( \frac{p_j^2}{2m_j}+ \frac{m_j\omega_j^2 x_j^2}{2}  \right), \label{Henv}
\end{eqnarray}
is given by the term
\begin{eqnarray}
H_{e- {\rm ph}} = - \sum_{jk} m_j\omega_j^2 x_{jk} x_j n_k, \label{Heph}
\end{eqnarray}
where $x_j$  and $p_j$ are the position and  momentum  of the $j$th oscillator having an effective mass $m_j$ and a frequency $\omega_j$. The
coefficients $x_{jk}$ define the strength of the coupling between the electron subsystem and the environment.

The contribution of the energy-quenching mechanisms responsible for the recombination processes in the system is given by the Hamiltonian
\begin{equation}
H_{\rm quen} = - \sum_l (q_l^\dagger a_{l^{*}}^\dagger a_l + q_l a_l^\dagger a_{l^{*}} ) . \label{Hquen}
\end{equation}
For the sake of simplicity, we include the radiation damping of the excited states into the energy-quenching operator $q_l$. The first
term in the Hermitian Hamiltonian $H_{\rm quen}$ is related to the excitation of the $l-$chromophore by the quenching bath, whereas the second term
corresponds to the reverse process, namely, to the absorption of chromophore energy by the bath. Both processes are necessary to provide correct
conditions for the thermodynamic equilibrium between the system and the bath.

The total Hamiltonian of the system is
\begin{eqnarray}
H = H_0 + H_{e- {\rm ph}}  + H_{\rm env} + H_{\rm quen}. \label{Htot1}
\end{eqnarray}
We omit here the Hamiltonian of the quenching (radiation) heat bath.

\subsection{Diagonalization of  $H_0$}
We choose 160 basis states $|M\rangle$ of the complex including a  vacuum state, where all chromophores are in the ground state and the F site
is empty. We diagonalize the Hamiltonian $H_0$  (\ref{H0}) to consider the case where the excitonic coupling between chromophores, described by
coefficients $V_{lm},$ and the porphyrin-fullerene tunneling, which is determined by amplitudes $\Delta_{\sigma \sigma'}$, cannot be analyzed
within perturbation theory. In the new basis, $|\mu\rangle = \sum_M | M \rangle \langle M|\mu \rangle$, the Hamiltonian $H_0$ is diagonal with
the energy spectrum \{$ E_{\mu}$\}, so that the total Hamiltonian of the system $H$ has the form
\begin{eqnarray}
H = \sum_{\mu} E_{\mu}| \mu\rangle \langle \mu | - \sum_{\mu \nu} {\cal A}_{\mu \nu} | \mu\rangle \langle \nu | + H_{\rm env}. \label{Htot2}
\end{eqnarray}
Here
\begin{equation} {\cal A}_{\mu\nu} = Q_{\mu \nu} + q_{\mu \nu} \label{calA}
\end{equation}
 is the combined operator for both heat baths with fluctuating in time variables
\begin{eqnarray}
Q_{\mu \nu} = \sum_{j} m_j\omega_j^2 x_j [  x_{jF}\langle \mu|n_F|\nu\rangle  + \sum_{k}  (x_{jk}\langle \mu|n_k|\nu\rangle + x_{jk^*}\langle
\mu|n_{k^*}|\nu\rangle ) ], \label{Qmunu1}
\end{eqnarray}
and
\begin{equation}
q_{\mu\nu} = \sum_l \,\langle \mu |a^{\dag}_{l}a_{l^*} |\nu \rangle \, q_l  + H.c. \label{qMuNu}
\end{equation}
To distinguish the processes of energy transfer, where the number of electrons on each chromophore remains constant, from the processes of
charge transfer, where the total population of the site changes, we introduce the following operators
\begin{eqnarray}
S_{l} = n_{l}+n_{l^*},\;\;M_{l} = n_{l}-n_{l^*}, \label{SM}
\end{eqnarray}
together with coefficients
\begin{equation}
\bar{x}_{jl} = \frac{x_{jl} + x_{jl^*}}{2}, \;\;\;\;\;\; \tilde{x}_{jl} = \frac{x_{jl} - x_{jl^*}}{2}.\label{xjl}
\end{equation}
Thus, the environment operator $Q_{\mu \nu}$ can be rewritten as
\begin{eqnarray}
Q_{\mu \nu} &=& \sum_{j} m_j\omega_j^2 x_j \Lambda^{\mu\nu}_j \label{Qmunu2}
\end{eqnarray}
with
\begin{eqnarray}
\Lambda^{\mu\nu}_j = \sum_{l}\left\{ \bar{x}_{jl}\langle \mu|S_l|\nu\rangle +  \tilde{x}_{j{l}}\langle \mu|M_{l}|\nu\rangle \right\} +
x_{jF}\langle \mu|n_F|\nu\rangle \label{LambdaMuNu}
\end{eqnarray}

\subsection{Non-Markovian equations for the system operators}
An arbitrary electron operator $W$ can be expressed in terms of the basic operators $\rho_{\mu\nu}~=~|\mu\rangle\langle \nu|$; with $ W =
\sum_{\mu\nu} W_{\mu\nu}\, \rho_{\mu\nu}, $ and $W_{\mu\nu} = \langle \mu| W |\nu \rangle. $ The operator $\rho_{\mu\nu}$ denotes a matrix with
zero elements, with the exception of the single element at the crossing of the $\mu-$row and the $\nu-$column. The matrix elements $W_{\mu\nu}$
of any electron operator can be easily calculated (see, e.g., Eqs.~(S10) and (S11) in the Supporting Information for Ref.~\cite{SmirnovJPC09}).
For example, an electron localized in a two-well potential \cite{Leggett87}, with the right and left states $|1\rangle$ and $ |2\rangle$, is
described by the Pauli matrices $\{\sigma_x, \sigma_y, \sigma_z \}:\; \sigma_z = |1\rangle \langle 1| - |2\rangle \langle 2|,\, \sigma_x =
|1\rangle \langle 2| + |2\rangle \langle 1|, \,$ and $ \sigma_y = i (|2\rangle \langle 1| - |1\rangle \langle 2|)$, which are expressed in terms
of the basic operators $|\mu\rangle \langle \nu|$ with  $\mu,\,\nu =1,2$.

In the Heisenberg picture, the operator $W$ evolves in time according to the equation: $i \left(\partial W/\partial t \right)= [W, H]_{-}\;.$ This evolution can be
described with the time-evolving operators, $\rho_{\mu\nu}(t)~=~(|\mu\rangle\langle \nu|)(t)$, which satisfy the Heisenberg equation:
\begin{equation}
i \frac{\partial {\rho}_{\mu\nu}}{\partial t} = [\rho_{\mu\nu}, H ]_-\; = \;-\, \omega_{\mu\nu} \rho_{\mu\nu} - \sum_{\alpha} ( {\cal A}_{\nu\alpha} \rho_{\mu\alpha} -
{\cal A}_{\alpha \mu} \rho_{\alpha \nu} ), \label{Heisenberg}
\end{equation}
where $\omega_{\mu\nu} = E_{\mu} - E_{\nu},$ and the heat bath operator ${\cal A}_{\mu\nu}$ is defined in Eq.~(\ref{calA}). Here, we use the
fact that the Hamiltonian $H$ Eq.~(\ref{Htot2}) is also expressed in terms of the operators $\rho_{\mu\nu}$ taken at the same  moment of time
$t$. For two of these operators, $\rho_{\mu\nu}(t)$ and $\rho_{\alpha \beta}(t)$, we have simple multiplication rules: $ \rho_{\mu\nu}
\rho_{\alpha\beta} = \delta_{\nu\alpha} \rho_{\mu\beta}.$ These rules allow to calculate commutators of basic operators taken at the same moment
of time. We note that at the initial moment of time the operator, $\rho_{\mu\nu}(0)~\equiv~|\mu\rangle \langle \nu| $, is represented by the
above-mentioned zero matrix with a single unit at the $\mu$-$\nu$ intersection. The matrix elements of the electron operators in
Eqs.~(\ref{qMuNu},\ref{LambdaMuNu}) are taken over the time-independent eigenstates of the Hamiltonian $H_0$. The bath operators ${\cal
A}_{\mu\nu}$ fluctuate in time since they depend on the environmental variables, $\{x_j(t)\}$, and on the variables $\{q_l(t)\}$ of the
quenching bath.

It is known that the dissipative evolution of the two-state system  can be described by the Heisenberg equations for the Pauli matrices
$\{\sigma_x, \sigma_y, \sigma_z \} $ with the spin-boson Hamiltonian [see Eq.~(1.4) in Ref.~\cite{Leggett87}], which includes environmental
degrees of freedom. The artificial photosynthetic complex analyzed in the present paper has 160 states. A dissipative evolution of this complex
is described by the Hamiltonian $H$ in Eq.~(\ref{Htot2}), written in terms of the Heisenberg operators $\rho_{\mu\nu}(t)~=~(|\mu\rangle\langle
\nu|)(t)$ taken at the moment of time $t$. Instead of the time-dependent Pauli matrices, the time evolution of the two-state dissipative system
can be described by the basic operators $|1\rangle\langle 1|,\, |1\rangle\langle 2|,\,|2\rangle\langle 1|,\,|2\rangle\langle 2|,$ evolving in
time. In a similar manner, the evolution of the multi-state photosynthetic complex is described by the set of the time-dependent Heisenberg
operators $\rho_{\mu\nu}(t)$, which obey the equation (\ref{Heisenberg}). As its spin-boson counterpart, the Hamiltonian $H$ in
Eq.~(\ref{Htot2}) contains the Hamiltonian, $H_{\rm env}$, of the heat bath as well as the system-bath interaction terms. Here, we generalize
the spin-boson model from the case of two states to the case of 160 states. With a knowledge of the operators $\rho_{\mu\nu}(t)$, it is possible
to find the time evolution of any Heisenberg operator of the system. Only at the initial moment of time, $t = 0$, the operators
$\rho_{\mu\nu}(0)$ form the basis of the Liouville space. Note that we work in the Heisenberg representation, without using the description
based on the von Neumann equations for the density matrix.

To obtain functions that can be measured in experiments, we have to average the operator $\rho_{\mu\nu}(t)$ and the equation (\ref{Heisenberg})
over the initial state $|\Psi^0\rangle$ of the electron subsystem as well as over the Gaussian distribution, $\rho_T = \exp(- H_{\rm
bath}^{(0)}/T),$ of the equilibrium bath, $\langle \ldots \rangle_T,$ with temperature $T$ and with a free Hamiltonian $H_{\rm bath}^{(0)}$,
which is comprised of the free environment Hamiltonian and the free Hamiltonian of the quenching bath. The notation $\langle \ldots \rangle$
means double averaging:
\begin{equation}
\langle \ldots \rangle  = \langle \langle \Psi^0|\ldots | \Psi^0 \rangle \rangle_T. \label{Av}
\end{equation}
The quantum-mechanical average value of the initial basic matrix,
$\langle \Psi^0|\rho_{\mu\nu}(0)|\Psi^0\rangle = \langle
\Psi^0|\mu\rangle\langle \nu|\Psi^0\rangle,$ is determined by the
product of amplitudes to find the electron subsystem at the initial
moment of time in the eigenstates $|\mu\rangle$ and $|\nu\rangle$ of
the Hamiltonian $H_0$.

A standard density matrix, $ \bar{\rho} = \{ \bar{\rho}_{\mu\nu}\}$, of the electron subsystem is a deterministic function which allows
to calculate the average value of an arbitrary operator $W$ with the formula:
\begin{equation}
\langle W(t) \rangle = Tr[\bar{\rho}(t)\, W] = \sum_{\mu\nu} W_{\mu\nu}\, \bar{\rho}_{\nu\mu}(t). \label{Wav1}
\end{equation}
The same average value can be written as $ \langle W(t) \rangle  = \sum_{\mu\nu} W_{\mu\nu}\, \langle \rho_{\mu\nu}(t)\rangle, $ which means
that the average matrix, $\langle \rho_{\mu\nu}(t)\rangle = \bar{\rho}_{\nu\mu}(t)$, has matrix elements related to the transposed density
matrix $\bar{\rho}(t)$.

 It should be emphasized that the time evolution of the heat-bath operators $\{ x_j,\,p_j\}$ and $ \{q_l\}$,  as well as their linear
combinations $Q_{\mu\nu},\, q_{\mu \nu}, $ and ${\cal A}_{\mu\nu} $, are determined by the total Hamiltonian $H$ in Eq.~(\ref{Htot2}). In the
absence of an interaction with the dynamical system (the electron-binding sites), the free-phonon operators $Q_{\mu\nu}^{(0)}$, as well as the
free operators of the other baths, $q_{\mu\nu}^{(0)}$, are described by Gaussian statistics \cite{Tanimura06}, as in the case of an environment
comprised of independent linear oscillators with the Hamiltonian $H_{\rm env}$ (\ref{Henv}).   Using the Gaussian property, Efremov and
coauthors \cite{Efremov81} derived non-Markovian Heisenberg-Langevin equations, without using perturbation theory, that assumes a weak
system-bath interaction. Recently, a similar non-perturbative approach has been developed by Ishizaki and Fleming in Ref.~\cite{IshizakiJCP09}.
Due to Gaussian properties of the free bath, the total operator ${\cal A}_{\mu\nu}$ of the combined dissipative environment is a linear
functional of the operators $\rho_{\mu\nu}$,
\begin{equation}
{\cal A}_{\mu\nu}(t) = {\cal A}_{\mu\nu}^{(0)}(t) + \sum_{\bar{\mu}\bar{\nu}} \int \langle i [ {\cal A}_{\mu\nu}^{(0)}(t), {\cal
A}_{\bar{\mu}\bar{\nu}}^{(0)}(t_1) ]_- \rangle \theta(t-t_1) \rho_{\bar{\mu}\bar{\nu}}(t_1), \label{Arho}
\end{equation}
where $\theta(\tau)$ is the Heaviside step function. We note that this expansion directly follows from the solution of the Heisenberg equations
for the positions $\{x_j\}$ and $\{q_l\}$ of the bath oscillators. It is shown in Ref.~\cite{Efremov81} that the average value of the free
operator ${\cal A}_{\mu\nu}^{(0)}(t)$ multiplied by an arbitrary operator ${\cal B}(t)$ is proportional to the functional derivative of the
operator ${\cal B}$ over the variable ${\cal A}_{\mu\nu}^{(0)}(t)$:
\begin{equation}
\langle {\cal A}_{\mu\nu}^{(0)}(t) {\cal B}(t) \rangle = \sum_{\bar{\mu}\bar{\nu}} \int dt_1 \;\langle {\cal A}_{\mu\nu}^{(0)}(t) {\cal
A}_{\bar{\mu}\bar{\nu}}^{(0)}(t_1) \rangle \times \left\langle \frac{\delta {\cal B}(t)}{ \delta {\cal A}_{\bar{\mu}\bar{\nu}}^{(0)}(t_1)
}\right\rangle, \label{Furutsu}
\end{equation}
with
\begin{equation}
\frac{\delta {\cal B}(t)}{ \delta {\cal A}_{\bar{\mu}\bar{\nu}}^{(0)}(t_1) } = i\; [ {\cal B}(t), \rho_{\bar{\mu}\bar{\nu}}(t_1) ]_-\;
\theta(t-t_1). \label{funcDer}
\end{equation}
Substituting Eqs.~(\ref{Arho},\ref{Furutsu},\ref{funcDer}) into Eq.~(\ref{Heisenberg}) we derive the exact non-Markovian equation for the
Heisenberg operators $\rho_{\mu\nu}$ of the dynamical system (chromomorphic sites + fullerene) interacting with a Gaussian heat bath,
\begin{eqnarray}
\langle \dot{\rho}_{\mu\nu}\rangle  - i \,\omega_{\mu\nu} \langle \rho_{\mu\nu}\rangle &=&
  \sum_{\alpha \bar{\mu}\bar{\nu}}
\int_0^t dt_1 \left\{ \langle   {\cal A}_{\bar{\mu}\bar{\nu}}^{(0)}(t_1) {\cal A}_{\nu\alpha}^{(0)}(t)\rangle
 \langle \rho_{\bar{\mu}\bar{\nu}}(t_1)\rho_{\mu\alpha}(t)\rangle \right.  \nonumber\\
  &-&
 \left.\langle {\cal A}_{\nu\alpha}^{(0)}(t) {\cal A}_{\bar{\mu}\bar{\nu}}^{(0)}(t_1) \rangle \langle \rho_{\mu\alpha}(t)
 \rho_{\bar{\mu}\bar{\nu}}(t_1)\rangle
   +  \langle {\cal A}_{\alpha\mu}^{(0)}(t) {\cal A}_{\bar{\mu}\bar{\nu}}^{(0)}(t_1) \rangle \langle
\rho_{\alpha\nu}(t) \rho_{\bar{\mu}\bar{\nu}}(t_1)\rangle
\right.  \nonumber\\
  &-&
 \left. \langle
 {\cal A}_{\bar{\mu}\bar{\nu}}^{(0)}(t_1) {\cal A}_{\alpha\mu}^{(0)}(t)\rangle
 \langle \rho_{\bar{\mu}\bar{\nu}}(t_1)\rho_{\alpha\nu}(t)\rangle \right.\}. \label{RhoEq1}
\end{eqnarray}
The time evolution of the average operator $\langle \rho_{\mu\nu}\rangle$ is determined by the second-order correlation functions of the system
operators as well as by the correlation functions of the free dissipative environment. Here we do not impose any restrictions on the spectrum of
the environment. It should be emphasized that the exact non-Markovian equation (\ref{RhoEq1}) goes far beyond the von Neumann equation, $i
\dot{\bar{\rho}} = [\bar{\rho}, H]_{-}, $ for the density matrix $\bar{\rho}$ of the electron subsystem.

\subsection{Beyond the system-bath perturbation theory.}
We assume that the coupling of the system to the quenching heat bath determined by the Hamiltonian $H_{\rm quen}$ (\ref{Hquen}) is weak enough
to be analyzed perturbatively. However, an interaction of the chromophores with the protein environment cannot be treated entirely within
perturbation theory since the reorganization energies are of the order of the intersite couplings. As in the theory of modified Redfield
equations \cite{Zhang98,YangFleming02}, the phonon operator $Q_{\mu\nu}$ in Eq.~(\ref{Qmunu2}) can be represented as a sum of diagonal $Q_{\mu}
= Q_{\mu \mu}$ and off-diagonal $\tilde{Q}_{\mu\nu}$ parts:
\begin{equation}
Q_{\mu\nu} = Q_{\mu} \delta_{\mu\nu} + (1 - \delta_{\mu\nu}) \tilde{Q}_{\mu\nu}.
\end{equation}
We derive equations for diagonal and off-diagonal elements of the matrix $\langle \rho_{\mu\nu}(t)\rangle$ (see Appendix B for details about the
derivation), where the interaction with the off-diagonal elements of the environment operators $\tilde{Q}_{\mu\nu}$  are considered within
perturbation theory, and the effects of the diagonal elements $Q_{\mu}$ are treated exactly.

The time dependence of the electron distribution  $\langle \rho_{\mu}\rangle$ (diagonal elements) over eigenstates of the Hamiltonian $H_0$ is
governed by the equation
\begin{equation}
\langle \dot{\rho}_{\mu}\rangle + \gamma_{\mu}\langle \rho_{\mu}\rangle = \sum_{\alpha} \gamma_{\mu\alpha} \langle \rho_{\alpha} \rangle,
\label{rhoMuTime}
\end{equation}
where the  relaxation matrix $\gamma_{\mu\alpha}$ contains a contribution, $\tilde{\gamma}_{\mu\alpha}$, from the non-diagonal environment
operators [see Eq.~(\ref{gamTilde})] as well as a contribution from the quenching processes, $\gamma_{\mu\alpha}^{\rm quen}$ [see
Eq.~(\ref{gamquen})],
\begin{equation}
\gamma_{\mu\alpha} = \tilde{\gamma}_{\mu\alpha} + \gamma_{\mu\alpha}^{\rm quen},
\end{equation}
with the total relaxation rate $\gamma_{\mu} = \sum_{\alpha} \gamma_{\alpha\mu}.$ The time evolution of the off-diagonal elements are given by
Eq.~(\ref{rhoMuNuTime}) in Appendix B.

Equations~(\ref{rhoMuTime},\ref{rhoMuNuTime}) allow us to determine the time evolution  of an average value for an arbitrary operator $W$ of the
system: $\langle W(t)\rangle = \sum_{\mu \nu} \langle \mu |W|\nu \rangle \langle \rho_{\mu\nu}(t) \rangle$.

\section{Energies and other parameters}
\subsection{Energy levels and electrochemical potentials} The energies of the excited states of chromophores BPEA, BDPY, and ZnPy, in the
BPF complex  are estimated from an average between the longest wavelength absorption band and the shortest wavelength emission band of the
chromophores. The average excited state energies of the chromophores BPEA, BDPY and ZnPy are 2610 meV, 2370 meV, and 2030 meV, respectively, if
we count from the corresponding ground energy levels \cite{gust2,gust3}.  Cyclic voltammetric studies \cite{gust3} of reduction potentials with
respect to the standard calomel electrode  show that the first reduction potential of the fullerene derivative, F, is about -- 0.62 V and the
first oxidation potential of ZnPy is about 0.75 V. From these data we calculate that the energy of the charge separated state ZnPy$^+-$F$^-$ is
about 1370 meV. This energy is a sum of the energy of an electron on  site F and a Coulomb interaction energy between a positive charge on ZnPy
and a negative charge on F. The Coulomb energy can be calculated with the formula $u = e^2/4\pi\epsilon_0 \epsilon r $, where $\epsilon_0$ is
the vacuum dielectric constant. The dielectric constant $\epsilon$ of 1,2 diflurobenzene (a solvent used in all experimental measurements of
Ref.~\cite{gust3}) is about 13.8. If the distance $r$ between porphyrin ZnPy and fullerene F is about 1 nm, the Coulomb interaction energy is
about 105 meV. Thus, the estimated energy of the electron on F can be of the order of 1475 meV.

\begin{tiny}
\begin{table}
\caption[] {This table presents the chosen values  of the excitonic couplings ($V$)  and reorganization energies for energy transfer ($\Lambda$)
of  the six antenna chromophores. We choose two sets of parameters, one set (denoted by I) corresponds to $V > \Lambda$ and the other set (II)
to the opposite limit $V < \Lambda$. The calculated values of the time constants using both sets of parameters agree with the experimental
values.}

\begin{tabular}{|c|c|c|c|c|}
  \hline
  Chromophores  & \begin{tabular}{c}  Set I\\Coupling (V)   \end{tabular}
  & \begin{tabular}{c}  Set I\\Reorganization\\ energy ($\Lambda$)   \end{tabular} &
    \begin{tabular}{c}  Set II\\Coupling (V)   \end{tabular} &
     \begin{tabular}{c}  Set II\\Reorganization\\ energy ($\Lambda$)
      \end{tabular} \\
  \hline
\begin{tabular}{c} BPEA$_a$ $\leftrightarrow$ BPEA$_b$,
\\BPEA$_b$ $\leftrightarrow$ BPEA$_a$ \end{tabular} &  50 meV &
\begin{tabular}{c} $\Lambda_{{\rm BPEA}_a} = 20$ meV \\ $\Lambda_{{\rm BPEA}_b}= 20$ meV
\end{tabular} &  30 meV &
\begin{tabular}{c} $\Lambda_{{\rm BPEA}_a} = 40$ meV \\ $\Lambda_{{\rm BPEA}_b}=
40$ meV \end{tabular}\\
   \hline
\begin{tabular}{c} BPEA$_a$ $\leftrightarrow$ BDPY$_a$,
\\BPEA$_b$ $\leftrightarrow$ BDPY$_b$ \end{tabular} &  30 meV &
\begin{tabular}{c} $\Lambda_{{\rm BDPY}_a} = 15$ meV \\ $\Lambda_{{\rm BDPY}_b}= 15$ meV
\end{tabular} &  17 meV &
\begin{tabular}{c} $\Lambda_{{\rm BDPY}_a} = 30$ meV \\ $\Lambda_{{\rm BDPY}_b}= 30$ meV
\end{tabular}\\
   \hline
\begin{tabular}{c} BDPY$_a$ $\leftrightarrow$ ZnPy$_a$,
\\BDPY$_b$ $\leftrightarrow$ ZnPy$_b$ \end{tabular} &  60 meV&
\begin{tabular}{c} $\Lambda_{{\rm ZnPy}_a} = 20$ meV \\ $\Lambda_{{\rm ZnPy}_b}= 20$ meV
\end{tabular} &  25 meV &
\begin{tabular}{c} $\Lambda_{{\rm ZnPy}_a} = 40$ meV \\ $\Lambda_{{\rm
ZnPy}_b} = 40$ meV \end{tabular}\\
   \hline
\begin{tabular}{c} BPEA$_a$ $\leftrightarrow$ ZnPy$_a$,
\\BPEA$_b$ $\leftrightarrow$ ZnPy$_b$ \end{tabular} &  50 meV&
- &  40 meV &-\\
   \hline
\begin{tabular}{c} BPEA$_b$ $\leftrightarrow$ ZnPy$_a$,
\\BPEA$_a$ $\leftrightarrow$ ZnPy$_b$ \end{tabular} &  60 meV & - &  40 meV &
- \\
   \hline
\end{tabular}
\end{table}
\end{tiny}

\subsection{Reorganization energies and coupling strengths } The reorganization energies for exciton and electron transfer
processes and electronic coupling strengths between the chromophores depend on the mutual distances and orientations of the components,
strengths of chemical bonds, solvent polarity and other structural details of the system. Precise values of these parameters are not available.
However, time constants for energy transfer between different chromophores in the BPF complex,  as well as rates for transitions of electrons
between the fullerene F and porphyrin chromophores ZnPy, have been reported in Ref.~\cite{gust3}. We fit the experimental values of these time
constants with the rates following from our equations with the goal of extracting reasonable values for the reorganization energies and the
electronic and excitonic couplings. In principle, many combinations of reorganization energies and coupling constants could be possible. For the
sake of simplicity, we consider two sets of parameters, for two limiting situations. One parameter set (denoted by I in Table I) corresponds to
a larger excitonic couplings, $V$, compared to the reorganization energies, $\Lambda$, whereas another set of parameters (denoted by II in Table
I) considers the opposite case: where the reorganization energies are larger than the excitonic couplings. These two sets of parameters are
presented in  Table I. In addition to the parameters listed in Table I, we consider the following values for the charge-transfer reorganization
energies (set I): $\lambda_F = 200$ meV, $\lambda_{lM} =100$~meV, and $\lambda_F = 230$ meV, $\lambda_{lM} =120$~meV (set II), where $ l = {\rm
ZnPy}_a, {\rm ZnPy}_b$. The values of the reorganization energies for energy-transfer processes are much smaller than those for charge transfer.

\begin{table}
\caption[] {This table presents a comparison between the calculated values of the time constants (using the parameters sets I and II) to the
experimental values reported in Ref.~\cite{gust3}. }

\begin{tabular}{|c|c|c|c|}
  \hline
  Process  &  $\tau$ (Set I) & $\tau$ (Set II) &
    $\tau$ (Experimental) \\
  \hline
   \begin{tabular}{c} BPEA$_a$ $\rightarrow$ BPEA$_b$,
    \\BPEA$_b$ $\rightarrow$ BPEA$_a$ \end{tabular} &  $\sim$ 0.4 ps &  $\sim$ 0.2 ps &  0.4 ps \\
   \hline
   \begin{tabular}{c} BPEA$_a$ $\rightarrow$ BDPY$_a$,
    \\BPEA$_b$ $\rightarrow$ BDPY$_b$ \end{tabular} &  $\sim$ 5 ps &  $\sim$ 5.4 ps &  5-13 ps \\
   \hline
   \begin{tabular}{c} BDPY$_a$ $\rightarrow$ ZnPy$_a$,
    \\BDPY$_b$ $\rightarrow$ ZnPy$_b$ \end{tabular} &  $\sim$ 5 ps &  $\sim$ 3.9 ps &  2-15 ps \\
   \hline
   \begin{tabular}{c} BPEA$_a$ $\rightarrow$ ZnPy$_a$,
    \\BPEA$_b$ $\rightarrow$ ZnPy$_b$ \end{tabular} &  $\sim$ 12 ps & $\sim$ 12 ps  &  7 ps \\
   \hline
  \begin{tabular}{c} BPEA$_b$ $\rightarrow$ ZnPy$_a$,
    \\BPEA$_a$ $\rightarrow$ ZnPy$_b$ \end{tabular} &  $\sim$ 10 ps &  $\sim$ 12 ps &  6 ps \\
   \hline
  \begin{tabular}{c} ZnPy$_b$ $\rightarrow$ F,
    \\ZnPy$_a$ $\rightarrow$ F \end{tabular} &  $\sim$ 3 ps &  $\sim$ 3 ps &  3 ps \\
  \hline
\end{tabular}
\end{table}
References \cite{gust3, gust4} reported a very fast electron transfer (with a time constant $\tau \sim$ 3 ps) between excited states of
zincporphyrins (ZnPy$_a$,ZnPy$_b$) and the fullerene derivative F. This fact indicates a good porphyrin-fullerene electronic coupling, which is
due to the short covalent linkage and close spatial arrangement of the components \cite{gust4}. Hereafter, we assume that the ZnPy-F tunneling
amplitudes $\Delta$ are about 100 meV (parameter set I) and 80 meV (parameter set II). These parameters provide a quite fast electron transfer,
despite of a significant energy gap between the ZnPy excited states and the fullerene energy level.

To describe recombination processes, we introduce a coupling of the $l$-th chromophore to a quenching heat-bath characterized for simplicity by
the Ohmic spectral density: $\chi''_l(\omega) = \alpha_l\, \omega$ with a dimensionless constant $\alpha_l$. We assume that the shifts of the
energy levels caused by the quenching bath are included into the renormalized parameters of the electron subsystem. The experimental values
\cite{gust3, gust4} of the lifetimes $\tau^e_l$ for excited states of chromophores BPEA, BDPY and ZnPy: $ \tau^e_{\rm BPEA} = 2.82 \;{\rm
ns},\;\tau^e_{\rm BDPY} = 0.26 \;{\rm ns}, \;$ and $\tau^e_{\rm ZnPY} = 0.45 \;{\rm ns},$ can be achieved  with the following set of coupling
constants: $\alpha_{\rm BPEA}  \sim 10^{-7} ,\;\alpha_{\rm BDPY} \sim  10^{-6} , \;\;{\rm and}\;\;\alpha_{\rm ZnPy} \sim 7 \times 10^{-7}.$

\section{Results and discussions}
Using Eqs.~(\ref{rhoMuNuTime},\ref{rhoMuTime}) and two sets of parameters discussed in Sec.~III, here we study electron and energy transfer
kinetics in the BPF complex  with special emphasis on the femtosecond time range, where the effects of quantum coherence can play an important
role. We consider both single- and double-exciton regimes.

\subsection{Evolution of a single exciton in the BPF complex  }

\begin{figure}[tp]
\centering
\includegraphics[width=12cm]{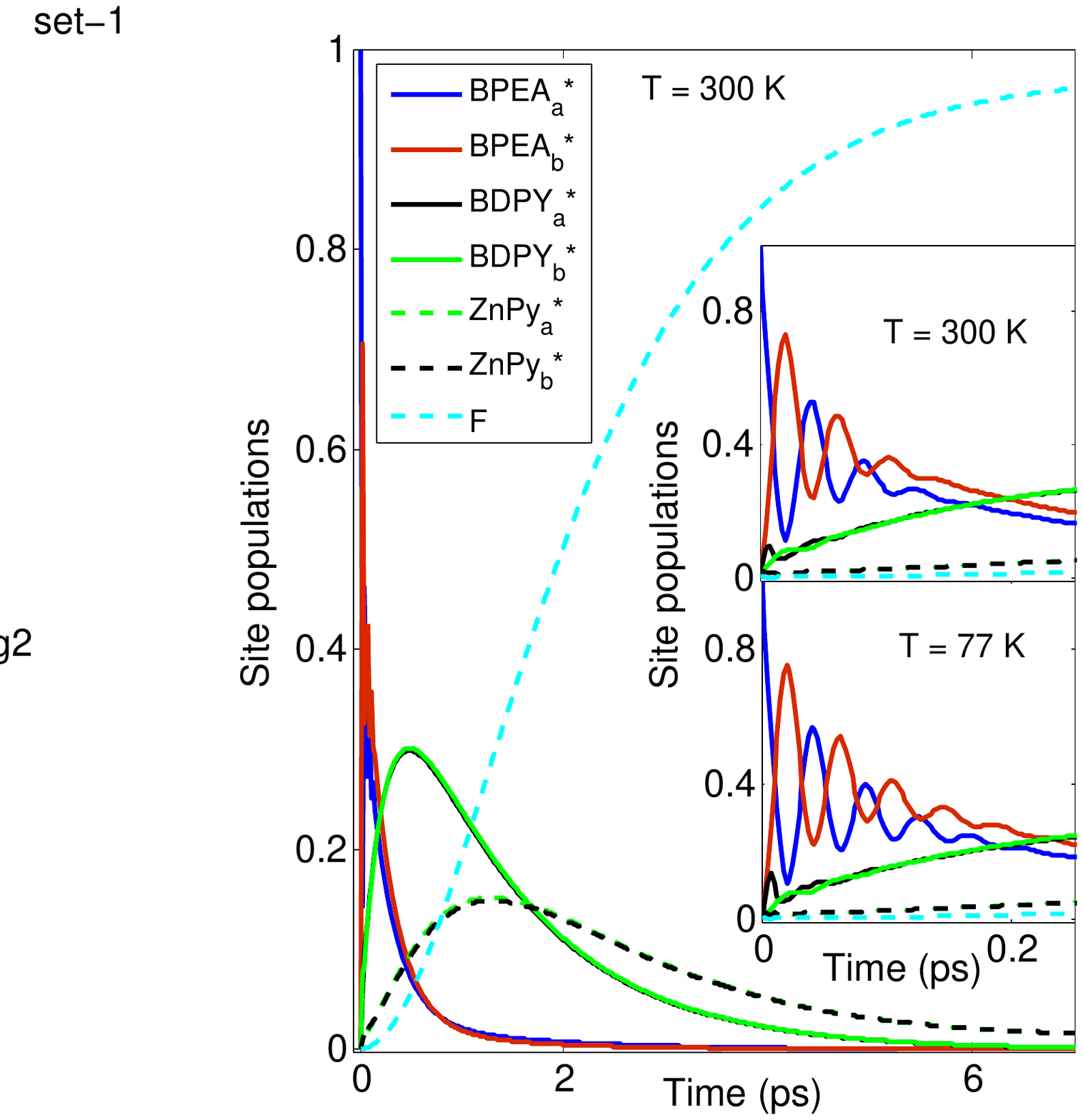}
\caption{(Color online) Site populations as a function of time for the parameter set I. The inset plots depict the features of site populations
for short times, at two different temperatures: $T$ = 300 K and 77 K. The site populations of the BPEA moieties oscillate with a considerably
large amplitude, while the oscillations of the other site populations are hardly observable. }\end{figure}

In Fig.~2 we show the time evolution of the excited states populations provided that only the   BPEA$_a$ chromophore  is excited at $t = 0$
(single-exciton regime). We use here the parameter set I, where excitonic couplings are larger than reorganization energies (see Sec.~III). The
process starts with quantum beatings between the resonant BPEA$_a$ and BPEA$_b$ chromophores, with a decoherence time of the order of 100 fs (at
$T$ = 300 K). In a few picoseconds, the excitation energy is subsequently transferred to the adjacent BDPY moieties and to the ZnPy
chromophores. Later on, an electron moves from the excited energy level of the porphyrins to the fullerene moiety; thus, producing a
charge-separated state, ZnPy$^+ -$F$^-$, with a quantum yield 95\%, which is in agreement with experimental results \cite{gust2}.
\begin{figure}[tp]
\centering
\includegraphics[width=12cm]{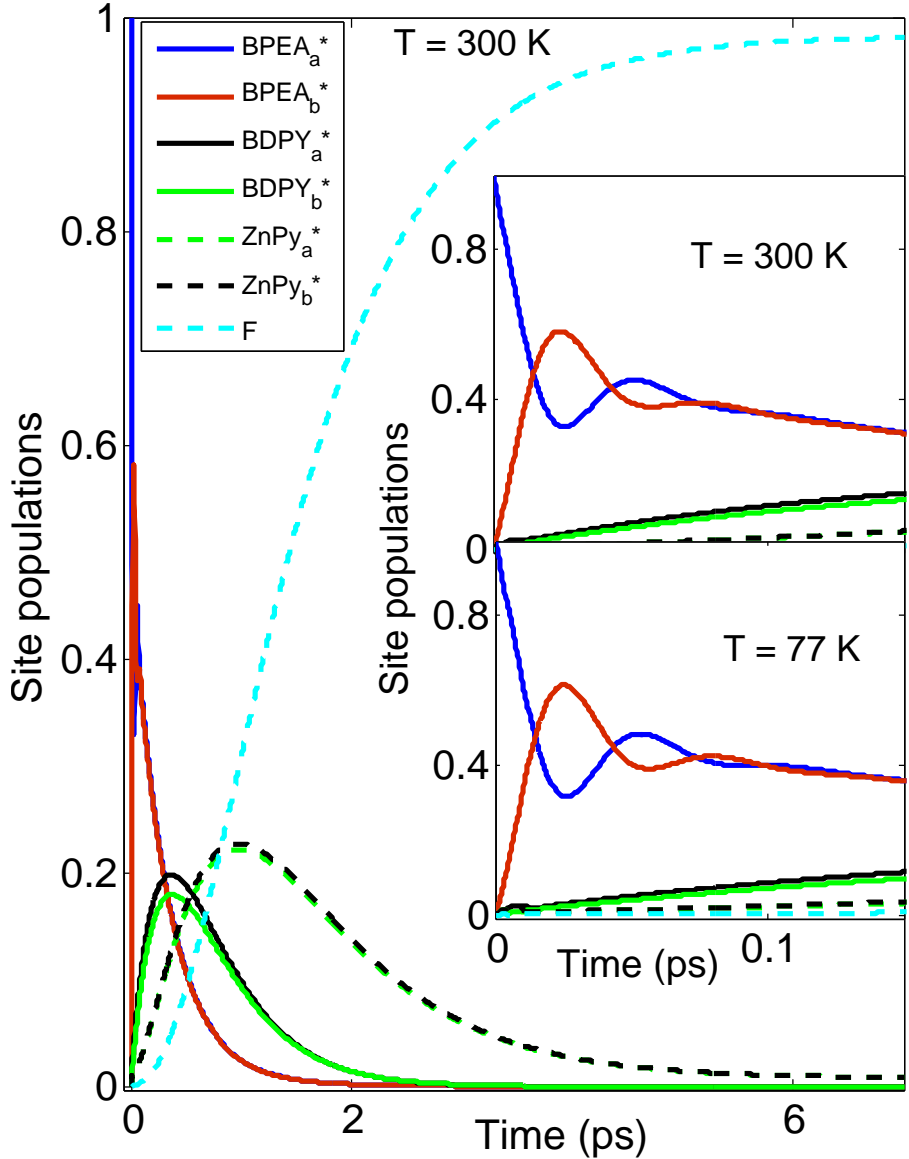}
\caption{(Color online) This figure presents site populations as a function of time for the parameter set II. The inset plots show
 the site populations for short times, at two
different temperatures: $T$ = 300 K and 77 K. The amplitudes of the site-population oscillations are much smaller and die out earlier, compared
to Fig.~2. This figure indicates that even for $\Lambda > V$, the energy transfer between BPEA chromophores is dominated by wave-like coherent
motion.}
\end{figure}
It is evident from Fig.~2 that excited state populations of the BDPY chromophores oscillate with much lower amplitudes and die out within a very
short time, $t < 10$ fs, at both temperatures: $T$ = 300 K and 77 K. The populations of the other sites of the BPF complex do not exhibit any
oscillatory behavior. This can be ascribed to incoherent hopping becoming dominant because of significant energy mismatch between these
chromophores.

\begin{figure}[tp]
\centering
\includegraphics[width=12cm]{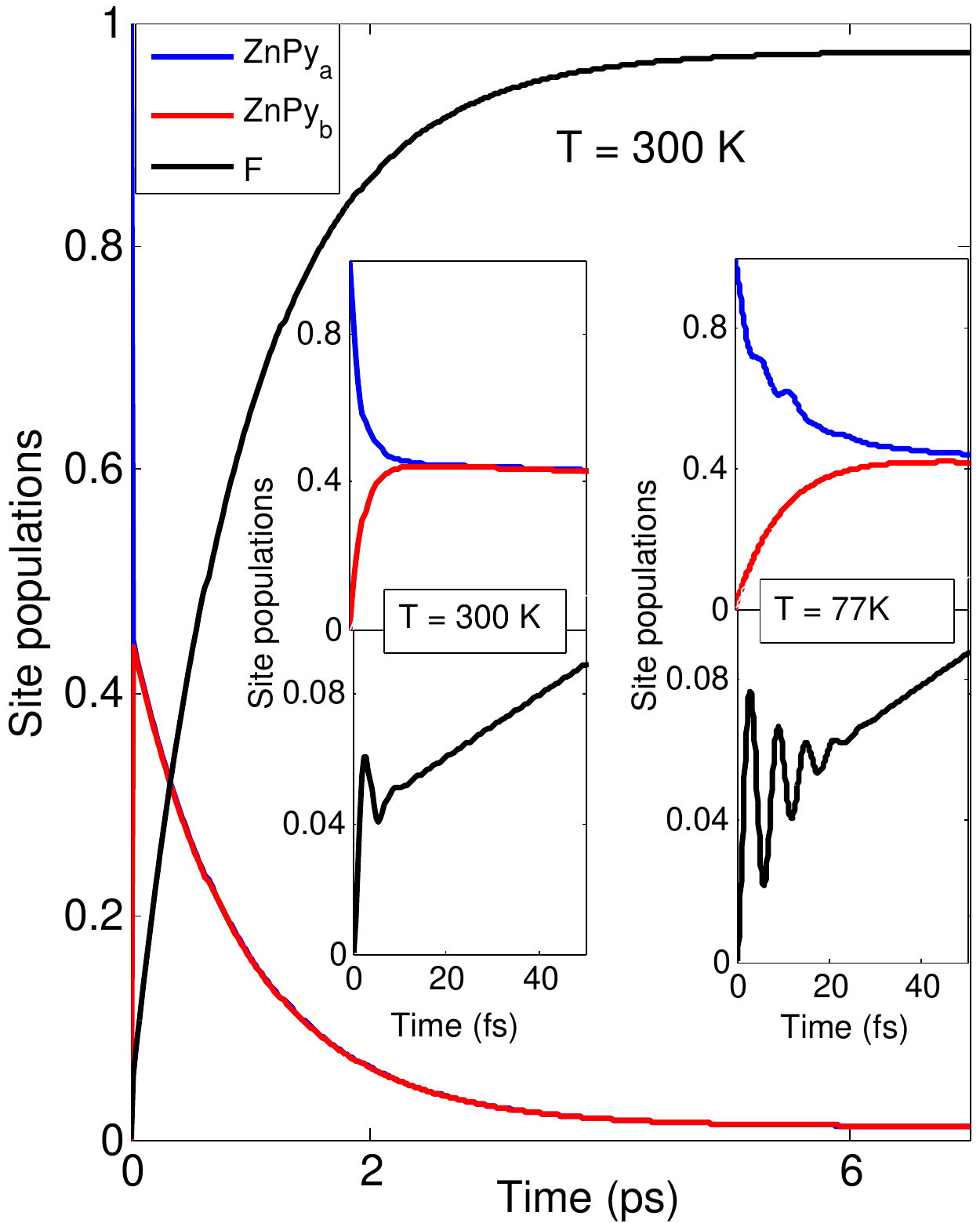}
\caption{(Color online) Site populations as a function of time for the parameter set I, when the ZnPy$_a$ chromophore  is in the excited state
and all the other chromophores are in the ground state at $t=0$. The inset plots depict the site populations at short times  for two
temperatures: $T$ = 300 and 77. Lowering the temperature enhances the oscillations of the charge density on the fullerene moiety. Despite the
huge energy difference between ZnPy$^*-$F and ZnPy$^+-$F$^-$, the charge of the fullerene site exhibits oscillatory behavior for short times,
specially at lower temperatures.}
\end{figure}

Figure 3 shows the time-dependence of the excited state populations of chromophores for the parameter set II, where the reorganization energies
are larger than the excitonic couplings between chromophores. At $t=0$ the BPEA$_a$ chromophore is excited (single-exciton regime). Then, after
a few picoseconds, the charge-separated state is formed with a quantum yield of the order of 97\%. However, owing to a stronger
system-environment coupling, quantum beats between the BPEA$_a$  and BPEA$_b$ chromophores have a lower amplitude and shorter decoherence time
($\sim$50 fs) than in the previous case when we use the parameter set I. We note that no quantum oscillations of the fullerene population (site
F) are visible in Figs.~2 and 3.

 No significant oscillations of the site populations were observed (not shown here) when
the BDPY chromophores were initially (at $t=0$) excited. In this case, due to the considerable energy gaps between the BDPY and the adjacent BPEA  and ZnPy
chromophores, incoherent hopping dominates over the coherent transfer of excitons. Furthermore, the structure of the BPF complex \cite{gust1,gust4}
does not allow direct energy transfer between two BDPY chromophores.

\begin{figure}[tp]
\centering
\includegraphics[width=9cm]{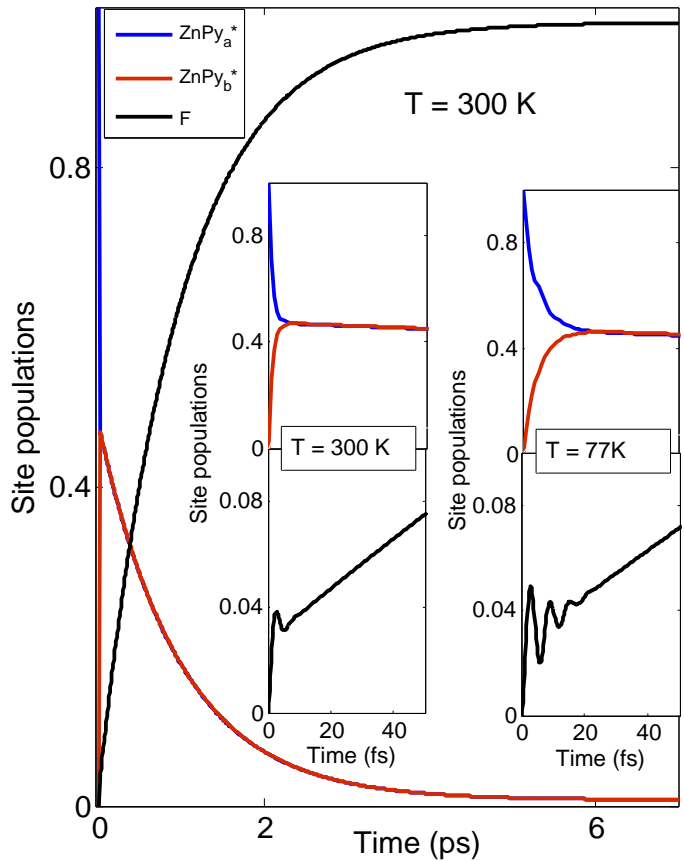}
\caption{(Color online) Time evolution of the site populations for the parameter set II, starting with an exciton on the chromophore ZnPy$_a$ at
$t=0$. The inset plots depict the features of the site populations for a shorter time regime and at two temperatures: $T$ = 300 K and 77 K.
Lowering the temperature enhances oscillations of the charge density on the fullerene derivative. These results indicate that the population of
the site F oscillates for short times, even for $\Lambda > V$. These oscillations are more pronounced at lower temperatures.}
\end{figure}
Figures 4 and 5 demonstrate charge- and energy-transfer dynamics for two parameter sets, I and II, for the case when one of the porphyrin
chromophores (ZnPy$_a$) is excited. Here we do not show the time evolution of the BPEA and BDPY chromophores since these moieties have higher
excitation energies than the ZnPy chromophore and they are not excited in the process. As evident from Figs.~4 and 5, the excited porphyrin
molecule rapidly transfers an electron to fullerene, thus, producing a charge-separated state ZnPy$^+-$F$^{-}$ with a quantum yield of about
98\%. The most important feature here is that the population and charge of the fullerene molecule oscillates in time due to a quantum
superposition of the porphyrin excited state and the state of an electron on the fullerene. The amplitude of these quantum beats is very small
and the decoherence time is quite short ($\sim$10 fs at T = 77~K). This fact can be explained by the significant energy mismatch between the
ZnPy$^*-$F and ZnPy$^+-$F$^-$ states as well as by the strong influence of the environment on the electron dynamics.

\subsection{Evolution of double excitons in the BPF complex }

\begin{figure}[tp]
\centering
\includegraphics[width=9cm]{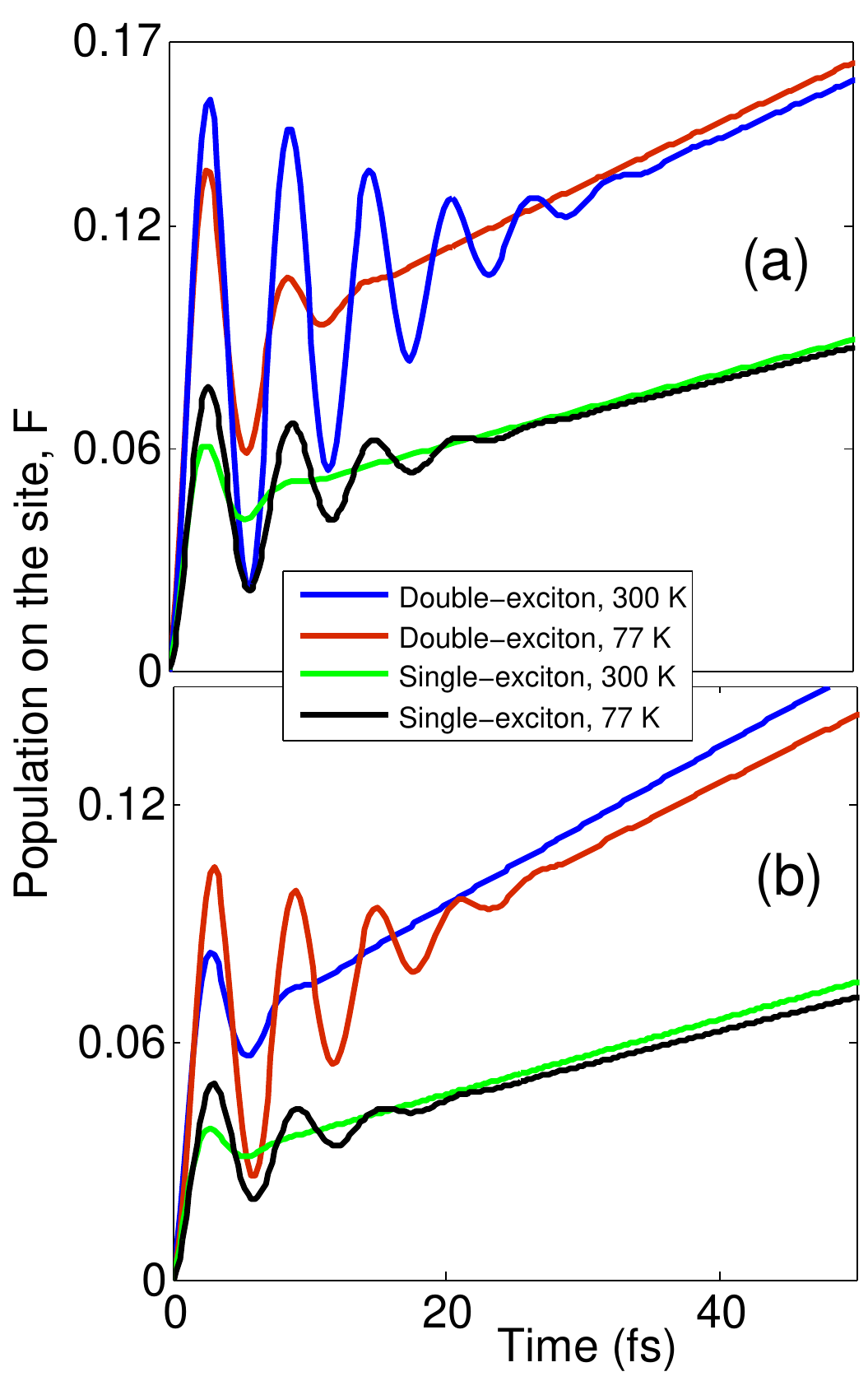}
\caption{(Color online) Time evolution of the populations on the site F, for both sets of parameters, I and II, comparing the double-exciton
case (the two ZnPy chromophores are excited) with the single-exciton case. (a) Time evolution of the populations on the site F for the parameter
set I. (b) Time evolution of the populations on the site F for the parameter set II. Note that the double-excitation significantly enhances the
amplitude of the charge oscillations at the fullerene site for both sets of parameters, either at low or high temperatures.}
\end{figure}

In the previous subsection, we consider a single exciton case with just one chromophore initially being in the upper energy state. Here we
analyze a situation where two porphyrin molecules (ZnPy$_a$ and ZnPy$_b$) are excited at $t=0$.
\begin{figure}[tp]
\centering
\includegraphics[width=9cm]{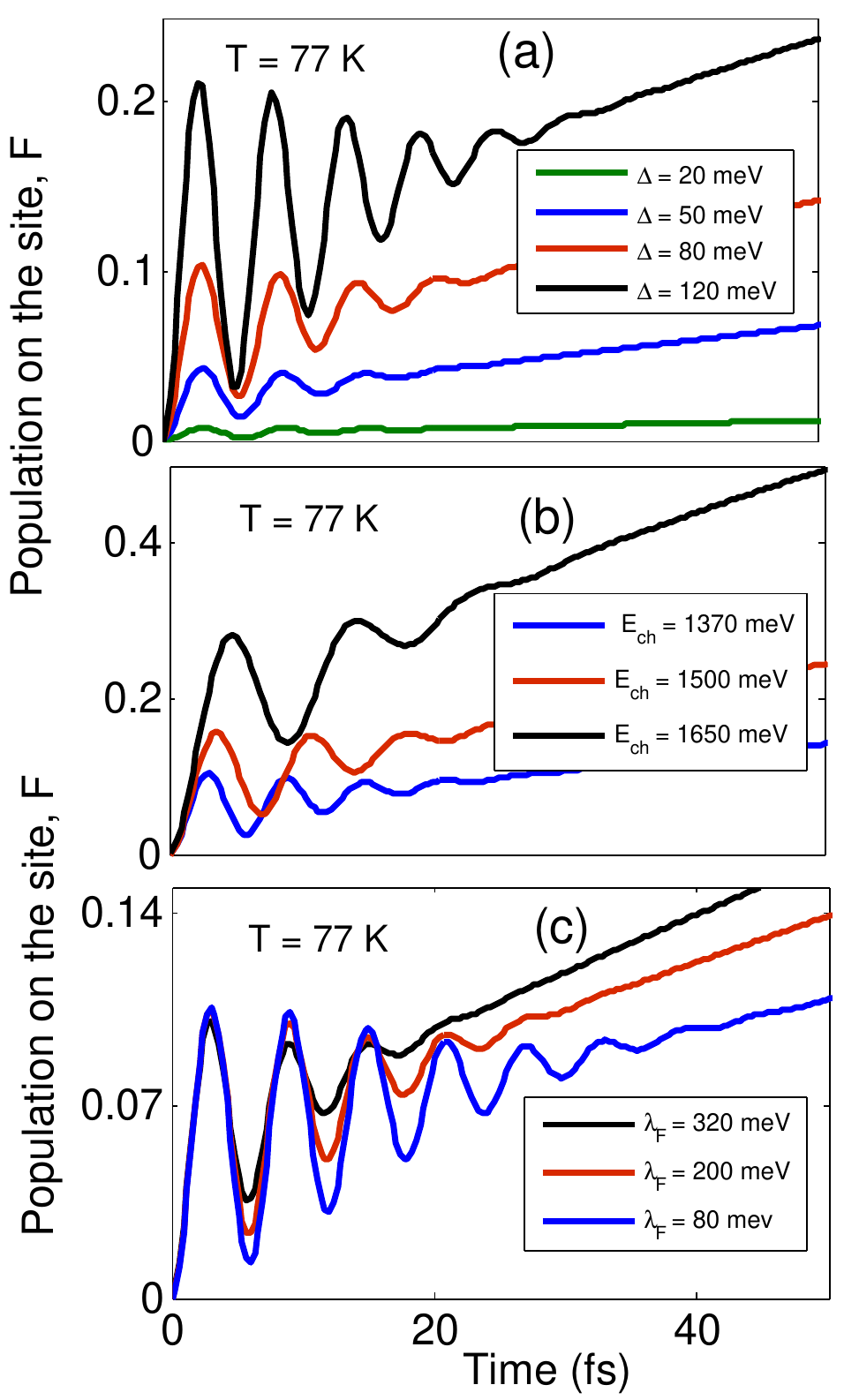}
\caption{(Color online) Time evolution of the population on the site F for the parameters set II when both  ZnPy chromophores are
 excited at $t=0$.  (a) Effects of the coupling $\Delta$ on the
time evolution of the populations on the site F. (b) Effects of the energy gap between an excited state of a ZnPy chromophore and the
charge-separated state, $E_{\rm ch}$, on the time evolution of populations on the site F. (c) Effects of the reorganization energy $\lambda$ on
the time evolution of populations on the site F. As can be seen from these plots, the contribution of wave-like coherent motion to
electron-transfer dynamics is significantly enhanced when strengthening the coupling between fullerene and porphyrin, lowering the energy gap
between the fullerene and porphyrin sites, and decreasing the reorganization energy.}
\end{figure}
Figures~6a and 6b show the coherent dynamics of the fullerene population (and the fullerene charge) for the parameter sets I (Fig.~6a) and II
(Fig.~6b) at two different temperatures, $T = 77$~K and $T = 300$~K. We also compare the double-exciton case with the previously analyzed
single-exciton case. It is apparent from Fig.~6, that the double excitation significantly enhances the amplitude of quantum oscillations of the
fullerene charge for both sets of parameters. As one might expect, the frequency of the quantum beatings and the decoherence time are not
affected by the number of excitons.

\subsection{Amplification of charge oscillations}
In the previous discussion we observed that lowering the temperature and the simultaneous excitation of both porphyrins significantly enhances
quantum oscillations of the fullerene charge. In this subsection we show that these oscillations can also be controlled by tuning the following
parameters:

\subsubsection{Electron tunneling amplitude $\Delta\,.$} The electronic coupling between the fullerene electron acceptor and zinc
porphyrins has a strong effect on the quantum oscillations of the fullerene charge. To explore this effect, in Fig.~7a we plot the electron
population of the fullerene as a function of time, for different values of the coupling $\Delta$. Figure~7a clearly shows that, with increasing
$\Delta$, the amplitude of the charge oscillations is significantly enhanced. This coupling can be increased by attaching the fullerene to
porphyrins with better ligands which form much stronger covalent bonds.

\subsubsection{Energy of the charge-separated state $E_{\rm ch}\,.$} The energy $E_{\rm ch} \sim$~1370~meV, of the charge separated
state, ZnPy$^+ -$F$^-$ is much lower than the energy of the zinc porphyrin excited state, $E_{{\rm ZnPy}^*} \sim$~2030~meV. It is evident from
Fig.~7b that increasing the energy $E_{\rm ch}$, which leads to a decrease of the porpyrin-fullerene energy mismatch, results in a pronounced
amplification of the quantum oscillations of the fullerene charge. The energy of the fullerene can be changed by placing nearby a charge
residue, electrostatically coupled to the fullerene.

\subsubsection{Reorganization energy $\lambda_F\,.$} In Fig.~7c we present the time evolution of the fullerene population for
different values of charge transfer reorganization energy $\lambda_F$. This parameter can be decreased by replacing the polar solvent with
another one which has a much lower polarity. As can be seen from Fig.~7c, the quantum oscillations of the fullerene charge survive much longer
times for smaller values of the reorganization energy, which correspond to weaker system-environment couplings. A similar effect is expected
when the porphyrin reorganization energy is changed.

\section{Conclusions}
We theoretically studied the energy and electron-transfer dynamics in a wheel-shaped artificial antenna-reaction center complex. This complex
\cite{gust3}, mimicking a natural photosystem, contains six chromophores (BPEA$_a$, BPEA$_b$, BDPY$_a$, BDPY$_b$, ZnPy$_a$, ZnPy$_b$) and an
electron acceptor (fullerene, F). Using methods of dissipative quantum mechanics  we derive and solve a set of equations for both the diagonal
and off-diagonal elements of the density matrix, which describe quantum coherent effects in energy and charge transfer. We consider two sets of
parameters, one corresponding to the case where the energy-transfer reorganization energy $\Lambda$ is less than the resonant coupling $V$
between the chromophores, $\Lambda < V$, and another regime where $\Lambda > V$. For these two sets of parameters we examine the electron and
exciton dynamics, with special emphasis on the short-time regime ($\sim$ femtoseconds). We demonstrate that, in agreement with experiments
performed in Ref. \cite{gust3}, the excitation energy of the BPEA antenna chromophores is efficiently funneled to porphyrins (ZnPy). The excited
ZnPy molecules rapidly donate an electron to the fullerene electron acceptor, thus creating a charge-separated state, ZnPy$^+-$F$^-$, with a
quantum yield of the order of 95\%. There is no observable difference in energy transduction efficiency for these two sets of
parameters. In the limit of strong interchromophoric coupling, coherent dynamics dominates over incoherent-hopping motion. In the single-exciton
regime, when one of the BPEA chromophores is initially excited, quantum beatings between two resonant BPEA chromophores occur with decoherence
times of the order of 100 fs. However, here the electron transfer process is dominated by incoherent hopping. For the case where one porphyrin
molecule is excited at the beginning, we obtain small quantum oscillations of the fullerene charge characterized by a short decay time scale
($\sim$ 10 fs). More pronounced quantum oscillations of the fullerene charge (with an amplitude $\sim$ 0.1 electron charge and decoherence time
of about 20 fs at $T$ = 77~K) are predicted for the double-exciton regime, when both porphyrin molecules are initially excited. We also show
that the contribution of wave-like coherent motion to electron-transfer dynamics could be enhanced by lowering the temperature, strengthening
the fullerene-porphyrin bonds, shrinking the energy gap between the zinc porphyrin and fullerene moieties (e.g., by attaching a charged residue
to the fullerene), as well as by  decreasing the reorganization energy (by tuning the solvent polarity).

\textbf{Acknowledgements.} FN acknowledges partial  support from the
Laboratory of Physical Sciences, National Security Agency, Army
Research Office, DARPA, Air Force Office of Scientific Research,
National Science Foundation grant No. 0726909, JSPS-RFBR contract
No.~09-02-92114, Grant-in-Aid for Scientific Research (S), MEXT
Kakenhi on Quantum Cybernetics, and Funding Program for Innovative
Research and Development  on Science and Technology (FIRST).

\begin{appendix}
\section{Coulomb interaction energies}

The Coulomb interactions between the electron states are,
\begin{eqnarray}\label{A1}
H_{\rm C} &=& -u_{\rm F}\left[(1-\bar{n}_{\rm ZnPy_a})n_{\rm F} + (1-\bar{n}_{\rm ZnPy_b})n_{\rm F}  \right] + u_{\rm Py} (1-\bar{n}_{\rm
ZnPy_a})(1-\bar{n}_{\rm ZnPy_b})\nonumber
\\
&+& u_{\rm ZnPy_a} n_{\rm ZnPy_a} n_{\rm ZnPy_a^*} + u_{\rm ZnPy_b} n_{\rm ZnPy_b} n_{\rm ZnPy_b^*},
\end{eqnarray}
 where,
 $$\bar{n}_{\rm ZnPy_a}=n_{\rm ZnPy_a} + n_{\rm
 ZnPy_a^*}\;\;\;\;\;
 {\rm and} \;\;\;\;\; \bar{n}_{\rm ZnPy_b}=n_{\rm ZnPy_b} + n_{\rm ZnPy_b^*}.$$
 The first term of (\ref{A1}) represents
 the electrostatic attraction (so the minus sign) between the
 positively charged ZnPy chromophores and the negatively-charged
 fullerene. The second term is due to the Coulomb repulsion (so the plus sign) between
 two ZnPy chromophores. The last two terms are the repulsive interaction energies when
 both the excited and ground states of the ZnPy chromophores are occupied by
 electrons. The coefficients $u_{\rm F}, u_{\rm Py},u_{\rm ZnPy_a},\;{\rm and} \;
 u_{\rm ZnPy_a}$ represent the magnitude of the electrostatic
 interactions and these are calculated using the Coulomb formula.
We have assumed that the empty ZnPy chromophores ($n_{\rm ZnPy} + n_{\rm ZnPy^*} = 0$) have positive charges and the acceptor state F becomes
negatively-charged when it is occupied by an electron.

\section{Derivation of equations for the matrix $\langle \rho_{\mu\nu}\rangle$}

Our derivation of the equations for the matrix $\langle \rho_{\mu\nu}\rangle$ is based on the exact solution for the operator $\rho_{\mu\nu} =
(|\mu\rangle\langle \nu|)(t)$ of the system influenced only by diagonal fluctuations of the bath. In this case the ``system + bath" Hamiltonian
has the form
\begin{equation}
H_{\rm diag} = \sum_{\mu} E_{\mu} |\mu\rangle \langle \mu| + \sum_j \left( \frac{p_j^2}{2 m_j} + \frac{m_j\omega_j^2 x_j^2}{2} \right) -
\sum_{\mu}\sum_j m_j\omega_j^2 \Lambda_j^{\mu} x_j |\mu\rangle \langle \mu|, \label{Hdiag}
\end{equation}
where $\Lambda_j^{\mu} = \Lambda_j^{\mu\mu}$ [see Eq.~(\ref{LambdaMuNu})]. The time evolution of the exciton operators $\rho_{\mu\nu}$ is
governed by the Heisenberg equation
\begin{equation}
i \dot{\rho}_{\mu\nu} = - \,\omega_{\mu\nu} \rho_{\mu\nu} + \sum_j m_j\omega_j^2 (\Lambda_j^{\mu} - \Lambda_j^{\nu}) x_j \rho_{\mu\nu}.
\label{RhoEq2}
\end{equation}
It is possible to verify that the solution of Eq.~(\ref{RhoEq2}) is given by the equation
\begin{eqnarray}
\rho_{\mu\nu}(t) = \exp[ i\Omega_{\mu\nu}(t-t_0)] \times \exp\left[ i \sum_j p_j(t)(\Lambda_j^{\mu} - \Lambda_j^{\nu}) \right] \times \nonumber\\
\exp\left[ - i \sum_j p_j(t_0)(\Lambda_j^{\mu} - \Lambda_j^{\nu}) \right] \rho_{\mu\nu}(t_0), \label{RhoEq3}
\end{eqnarray}
where
\begin{equation}
\Omega_{\mu\nu} = \omega_{\mu\nu} - \sum_j \frac{m_j\omega_j^2}{2} \left[ (\Lambda_j^{\mu})^2 - (\Lambda_j^{\nu})^2 \right], \label{OmegaMuNu}
\end{equation}
and $p_j$ is the Heisenberg operator of the dissipative environment. The evolution begins at time $t=t_0$. The diagonal operators $\rho_{\mu} =
\rho_{\mu\mu}$ are constant, $\rho_{\mu}(t) = \rho_{\mu}(t_0)$, in the presence of a strong interaction with the diagonal operators of the
protein environment.

For uncorrelated diagonal and off-diagonal environment operators, when $\langle Q^{(0)}_{\alpha}(t) \tilde{Q}^{(0)}_{\mu\nu}(t')\rangle = 0$,
the contribution of the environment to the non-Markovian equation (\ref{RhoEq1}) consists of two parts:
\begin{eqnarray}
\langle -i [ \rho_{\mu\nu}, H_{e- {\rm ph}} ]_{-}\rangle = \langle -i [ \rho_{\mu\nu}, H_{e- {\rm ph}} ^{\rm diag}]_{-}\rangle + \langle -i [
\rho_{\mu\nu}, H_{e- {\rm ph}} ^{\rm n-diag}]_{-}\rangle.
\end{eqnarray}
The diagonal elements, $Q_{\mu}$, of the environment contribute to the first part,
\begin{eqnarray}
\langle -i [ \rho_{\mu\nu}, H_{e- {\rm ph}} ^{\rm diag}]_{-}\rangle = \int_0^t dt_1 \langle (Q_{\mu}^{(0)} - Q_{\nu}^{(0)})(t)
Q_{\bar{\nu}}^{(0)}(t_1)\rangle \langle \rho_{\mu\nu}(t)\rho_{\bar{\nu}}(t_1)\rangle - \nonumber\\
\int_0^t dt_1 \langle Q_{\bar{\nu}}^{(0)}(t_1)(Q_{\mu}^{(0)} - Q_{\nu}^{(0)})(t) \rangle \langle \rho_{\bar{\nu}}(t_1) \rho_{\mu\nu}(t)\rangle,
\label{RhoQDiag}
\end{eqnarray}
whereas the second part is due to a contribution of the non-diagonal (abbreviated as n-diag in the super-index) operators, $\tilde{Q}_{\mu\nu}$,
\begin{eqnarray}
\langle -i [ \rho_{\mu\nu}, H_{e- {\rm ph}} ^{\rm n-diag}]_{-}\rangle = - \int_0^t dt_1 \langle
\tilde{Q}_{\nu\alpha}^{(0)}(t)\tilde{Q}_{\bar{\mu}\bar{\nu}}^{(0)}(t_1) \rangle \langle \rho_{\mu\alpha}(t)\rho_{\bar{\mu}\bar{\nu}}(t_1)\rangle
+ \nonumber\\ \int_0^t dt_1 \langle \tilde{Q}_{\bar{\mu}\bar{\nu}}^{(0)}(t_1)\tilde{Q}_{\nu\alpha}^{(0)}(t) \rangle \langle
\rho_{\bar{\mu}\bar{\nu}}(t_1)\rho_{\mu\alpha}(t)\rangle + \nonumber\\ \int_0^t dt_1 \langle
\tilde{Q}_{\alpha\mu}^{(0)}(t)\tilde{Q}_{\bar{\mu}\bar{\nu}}^{(0)}(t_1) \rangle \langle \rho_{\alpha\nu}(t)\rho_{\bar{\mu}\bar{\nu}}(t_1)\rangle
- \nonumber\\ \int_0^t dt_1 \langle \tilde{Q}_{\bar{\mu}\bar{\nu}}^{(0)}(t_1)\tilde{Q}_{\alpha\mu}^{(0)}(t) \rangle \langle
\rho_{\bar{\mu}\bar{\nu}}(t_1)\rho_{\alpha\nu}(t)\rangle. \label{RhoQnonDiag}
\end{eqnarray}
We note that the time evolution of the diagonal elements of the system operator, $\rho_{\mu} = \rho_{\mu\mu},$ is determined by the non-diagonal
operators $\tilde{Q}_{\mu\nu}$ as well as by quenching terms. Strong diagonal fluctuations of the environment have no effect on the evolution of
the diagonal elements of the matrix. Thus, in Eq.~(\ref{RhoQDiag}) we assume that $\rho_{\bar{\nu}}(t_1) = \rho_{\bar{\nu}}(t),$ so that
Eq.~(\ref{RhoQDiag}) can be rewritten as
\begin{eqnarray}
\langle -i [ \rho_{\mu\nu}, H_{e- {\rm ph}} ^{\rm diag}]_{-}\rangle = - (\Gamma_{\mu\nu}^{\rm diag}+i\delta\Omega_{\mu\nu}^{\rm diag})(t)
\langle \rho_{\mu\nu}(t)\rangle, \label{rhoMuNuDiag}
\end{eqnarray}
where the time-dependent rate, $\Gamma_{\mu\nu}^{\rm diag}(t)$, and the frequency shift, $ \delta\Omega_{\mu\nu}^{\rm diag}$, can be found from
the following expression
\begin{eqnarray}
\Gamma_{\mu\nu}^{\rm diag}(t)+i\delta\Omega_{\mu\nu}^{\rm diag}(t) = \int_0^t dt_1 \left\{ \left\langle (Q_{\mu}^{(0)} - Q_{\nu}^{(0)})(t)
Q_{\nu}^{(0)}(t_1)\right\rangle -  \left\langle Q_{\mu}^{(0)}(t_1)(Q_{\mu}^{(0)} - Q_{\nu}^{(0)})(t) \right\rangle \right\}.
\end{eqnarray}
The  rate $\Gamma_{\mu\nu}^{\rm diag}(t)$ determines the fast decay of quantum coherence in our system. For an environment composed of
independent oscillators we obtain
\begin{eqnarray}
\langle (Q_{\mu}^{(0)} - Q_{\nu}^{(0)})(t) Q_{\nu}^{(0)}(t_1)\rangle -  \langle Q_{\mu}^{(0)}(t_1)(Q_{\mu}^{(0)} - Q_{\nu}^{(0)})(t) \rangle =
\nonumber\\ - \sum_j \frac{m_j\omega_j^3}{2} (\Lambda_j^{\mu} - \Lambda_j^{\nu})^2 \coth\left(\frac{\omega_j}{2T}\right) \cos \omega_j(t-t_1) -
\nonumber\\ i \sum_j \frac{m_j\omega_j^3}{2} \left[(\Lambda_j^{\mu})^2 - (\Lambda_j^{\nu})^2\right]  \sin \omega_j(t-t_1). \label{Q0}
\end{eqnarray}
The fluctuations of the diagonal operators of the environment can be described by the set of spectral functions,
\begin{eqnarray}
J_{\mu}(\omega) = \sum_j \frac{m_j\omega_j^3}{2} (\Lambda_j^{\mu})^2 \delta(\omega - \omega_j), \nonumber\\
\bar{J}_{\mu\nu}(\omega) = \sum_j \frac{m_j\omega_j^3}{2} (\Lambda_j^{\mu} - \Lambda_j^{\nu})^2 \delta(\omega - \omega_j),
\end{eqnarray}
together with the corresponding reorganization energies,
\begin{eqnarray}
\lambda_{\mu} = \int_0^{\infty}\frac{d\omega}{\omega} J_{\mu}(\omega) = \sum_j \frac{m_j\omega_j^2}{2} (\Lambda_j^{\mu})^2, \nonumber\\
\bar{\lambda}_{\mu\nu} = \int_0^{\infty}\frac{d\omega}{\omega} \bar{J}_{\mu\nu}(\omega) = \sum_j \frac{m_j\omega_j^2}{2} (\Lambda_j^{\mu} -
\Lambda_j^{\nu})^2. \label{barlambda}
\end{eqnarray}
We also introduce a spectral function, $\tilde{J}_{\mu\nu}(\omega)$, which characterizes the non-diagonal ($\mu \neq \nu$) environment
fluctuations,
\begin{equation}
\tilde{J}_{\mu\nu}(\omega) = \sum_j \frac{m_j\omega_j^3}{2} |\tilde{\Lambda}_j^{\mu\nu}|^2 \delta(\omega - \omega_j),
\end{equation}
where $\tilde{\Lambda}_j^{\mu\nu} = \Lambda_j^{\mu\nu}$ (\ref{LambdaMuNu}) taken at $\mu\neq \nu$. With Eq.~(\ref{Q0}) we calculate the
contributions of the diagonal environment fluctuations into the decoherence rate and the frequency shift of the off-diagonal elements of the
system matrix $\langle \rho_{\mu\nu} \rangle $ in (\ref{rhoMuNuDiag}),
\begin{eqnarray}
\Gamma_{\mu\nu}^{\rm diag}(t) = \int_0^{\infty} \frac{d\omega}{\omega} \bar{J}_{\mu\nu}(\omega) \coth\left( \frac{\omega}{2T}\right) \sin \omega
t,
\nonumber\\
\delta\Omega_{\mu\nu}^{\rm diag}(t) = \int_0^{\infty} \frac{d\omega}{\omega} [ J_{\mu}(\omega) - J_{\nu}(\omega) ] (1 - \cos\omega t ).
\label{GamOmMunu}
\end{eqnarray}

The contribution of the non-diagonal fluctuations of the environment to the evolution of the electron operators $\langle \rho_{\mu\nu}\rangle$
is defined by Eq.~(\ref{RhoQnonDiag}). To calculate the products of exciton variables taken at different moments of time, for example,
$\rho_{\mu\alpha}(t) \rho_{\bar{\mu}\bar{\nu}}(t_1)$, we use Eq.~(\ref{RhoEq3}), which describes the evolution of exciton operators in the
presence of strong coupling to the diagonal operators, $Q_{\mu}$, of the environment. We assume that the interaction with the non-diagonal
environment operators, $\tilde{Q}_{\mu\nu}$, is weak. With Eq.~(\ref{RhoEq3}) we express the operators at time $t_1$ in terms of operators taken
at time $t$:
\begin{eqnarray}
\rho_{\bar{\mu}\bar{\nu}}(t_1) = \exp\left[- i \Omega_{\bar{\mu}\bar{\nu}}\tau \right] \exp\left[i u_{\bar{\mu}\bar{\nu}}(\tau)\right]
\exp\left[-i v_{\bar{\mu}\bar{\nu}}(t,t_1)\right]
\rho_{\mu\nu}(t), \nonumber\\
\rho_{\bar{\mu}\bar{\nu}}(t_1) = \rho_{\mu\nu}(t) \exp\left[- i \Omega_{\bar{\mu}\bar{\nu}}\tau\right] \exp\left[-i
u_{\bar{\mu}\bar{\nu}}(\tau)\right] \exp\left[-i v_{\bar{\mu}\bar{\nu}}(t,t_1)\right], \label{RhoTime}
\end{eqnarray}
where $\tau = t-t_1$, and
\begin{eqnarray}
u_{\mu\nu}(\tau) = \int_0^{\infty} \frac{d\omega}{\omega} \bar{J}_{\mu\nu}(\omega) \sin \omega \tau, \nonumber\\
v_{\mu\nu}(t,t_1) = \sum_j (\Lambda_j^{\mu} - \Lambda_j^{\nu}) [ p_j(t) - p_j(t_1)].
\end{eqnarray}
Here we assume that $p_j(t), p_j(t_1)$ are free-evolving momentum operators of the environment, which are described by Gaussian statistics with
a correlation function
\begin{equation}
\left \langle\frac{1}{2} \left[\; p_j(t),p_j(t_1)\right]_{+}\right\rangle = \frac{\hbar m_j\omega_j}{2} \coth\left(\frac{\hbar\omega_j}{2
T}\right) \cos \omega_j (t-t_1).
\end{equation}
 The operator function $v_{\mu\nu}(t,t_1)$ does not commute with the exciton matrix $\rho_{\mu\nu}(t)$, and, therefore, we
need two expressions for the operator $\rho_{\mu\nu}(t_1)$, which are distinguished by the order of the operators $\rho_{\mu\nu}(t)$ and
$\exp\left[-i v_{\mu\nu}(t,t_1)\right].$ For the average value of the operator $\exp\left[-i v_{\mu\nu}(t,t_1)\right]$ we obtain
\begin{eqnarray}
\langle\exp\left[-i v_{\mu\nu}(t,t_1)\right]\rangle = \exp\left\{ - \int_0^{\infty} \frac{d\omega}{\omega^2} \bar{J}_{\mu\nu}(\omega)
\coth\left(\frac{\hbar\omega}{2 T}\right) [ 1 - \cos \omega(t-t_1) ]\right\}.
\end{eqnarray}

Substituting Eqs.~(\ref{RhoTime}) to Eq.~(\ref{RhoQnonDiag}) and using the secular approximation we obtain a contribution of the non-diagonal
environment operators, $\tilde{Q}_{\mu\nu}$, to the evolution of diagonal exciton operators $\langle \rho_{\mu} \rangle$,
\begin{eqnarray}
\langle -i [ \rho_{\mu}, H_{e- {\rm ph}} ^{\rm n-diag}]_{-}\rangle = - \sum_{\alpha} \tilde{\gamma}_{\alpha\mu}(t) \langle \rho_{\mu} \rangle +
\sum_{\alpha} \tilde{\gamma}_{\mu\alpha}(t) \langle \rho_{\alpha} \rangle,
\end{eqnarray}
characterized by the following relaxation matrix,
\begin{eqnarray}
\tilde{\gamma}_{\mu\alpha}(t) = \int_0^t dt_1 \langle \tilde{Q}_{\alpha\mu}^{(0)}(t)\tilde{Q}_{\mu\alpha}^{(0)})(t_1) \rangle
e^{-i\Omega_{\mu\alpha}(t-t_1)} e^{ - i u_{\mu\alpha}(t-t_1)} \langle e^{-i v_{\mu\alpha}(t,t_1)}\rangle + \nonumber\\
\int_0^t dt_1 \langle \tilde{Q}_{\alpha\mu}^{(0)}(t_1)\tilde{Q}_{\mu\alpha}^{(0)})(t) \rangle e^{-i\Omega_{\alpha\mu}(t-t_1)} e^{ i
u_{\alpha\mu}(t-t_1)} \langle e^{-i v_{\alpha\mu}(t,t_1)}\rangle, \label{gamTime}
\end{eqnarray}
where
\begin{eqnarray}
\langle \tilde{Q}_{\alpha\mu}^{(0)}(t)\tilde{Q}_{\mu\alpha}^{(0)})(t_1) \rangle = (1/2) \int_0^{\infty} \tilde{J}_{\alpha\mu}(\omega)\times
\nonumber\\ \left\{ \left[ \coth\left(\frac{\omega}{2 T}\right) - 1 \right] e^{i\omega(t-t_1)} + \left[ \coth\left(\frac{\omega}{2 T}\right) + 1
\right] e^{-i\omega(t-t_1)}\right\}.
\end{eqnarray}
When the environment is at high temperatures ($ 2T \gg \omega$) and at low frequencies of the diagonal fluctuations ($\omega\tau \ll 1$) we
have: $$ u_{\mu\nu} (\tau) \simeq \bar{\lambda}_{\mu\nu}\tau,
$$ and $$\langle \exp[ - i v_{\mu\nu}(t,t_1) ] \rangle \simeq
\exp[ - \bar{\lambda}_{\mu\nu} T (t-t_1)^2 ].$$ With these assumptions the relaxation matrix has a simple form
\begin{eqnarray}
\tilde{\gamma}_{\mu\alpha} = \sqrt{\frac{\pi}{\bar{\lambda}_{\alpha\mu}}}
\int_0^{\infty} d\omega \,\tilde{J}_{\alpha\mu}(\omega)\,  n(\omega)\times \nonumber\\
\left\{ \exp\left[ - \frac{(\omega + \Omega_{\alpha \mu} - \bar{\lambda}_{\alpha\mu})^2}{ 4 \bar{\lambda}_{\alpha\mu} T} \right] + \exp\left(
\frac{\omega}{T}\right) \exp\left[ - \frac{(\omega - \Omega_{\alpha \mu} + \bar{\lambda}_{\alpha\mu})^2}{ 4 \bar{\lambda}_{\alpha\mu} T} \right]
\right\}, \label{gamTilde}
\end{eqnarray}
where $n(\omega) = [\exp(\omega/T) -1]^{-1}$ is the Bose distribution function at the temperature $T$. The moment of time $t$ in the expression
(\ref{gamTime}) for the relaxation matrix is usually higher than the effective retardation time, $\tau_c \sim (\bar{\lambda}_{\alpha \mu}
T)^{-1/2}$, of the integrand in Eq.~(\ref{gamTime}): $t\gg \tau_c$. Therefore, we assume that $t \simeq \infty$, so that
$\tilde{\gamma}_{\mu\alpha}(t) \simeq \tilde{\gamma}_{\mu\alpha}(\infty)
 = \tilde{\gamma}_{\mu\alpha}.$

It follows from Eq.~(\ref{RhoQnonDiag}) that  a contribution of the non-diagonal environment operators $\tilde{Q}_{\mu\nu}$ to the evolution of
the off-diagonal elements $\rho_{\mu\nu}$  is given by the formula
\begin{eqnarray}
\langle -i [ \rho_{\mu\nu}, H_{e- {\rm ph}} ^{\rm n-diag}]_{-}\rangle = - (\tilde{\Gamma}_{\mu\nu}+i\delta\tilde{\Omega}_{\mu\nu})(t) \langle
\rho_{\mu\nu}(t)\rangle,
\end{eqnarray}
where
\begin{eqnarray}
\tilde{\Gamma}_{\mu\nu}(t)+i\delta\tilde{\Omega}_{\mu\nu}(t) = \int_0^t dt_1 \langle
\tilde{Q}_{\nu\alpha}^{(0)}(t)\tilde{Q}_{\alpha\nu}^{(0)}(t_1) \rangle e^{-i\Omega_{\alpha\nu}(t-t_1)} e^{- i u_{\alpha\nu}(t-t_1)} \langle
e^{-i v_{\alpha\nu}(t,t_1)}\rangle + \nonumber\\
\int_0^t dt_1 \langle \tilde{Q}_{\mu\alpha}^{(0)}(t)\tilde{Q}_{\alpha\mu}^{(0)}(t_1) \rangle e^{-i\Omega_{\mu\alpha}(t-t_1)} e^{ i
u_{\mu\alpha}(t-t_1)} \langle e^{-i v_{\mu\alpha}(t,t_1)}\rangle.
\end{eqnarray}
A small frequency shift, $\delta\tilde{\Omega}_{\mu\nu},$ can be hereafter ignored. The dephasing rate, $\tilde{\Gamma}_{\mu\nu}$, has two
parts, $\tilde{\Gamma}_{\mu\nu} = \tilde{\Gamma}_{\mu} + \tilde{\Gamma}_{\nu}, $ where
\begin{eqnarray}
\tilde{\Gamma}_{\mu} = \frac{1}{2} \sum_{\alpha} \sqrt{\frac{\pi}{\bar{\lambda}_{\mu\alpha} T}} \int_0^{\infty} d\omega
\tilde{J}_{\mu\alpha}(\omega) n(\omega) \times \nonumber\\
\left\{ \exp\left[ - \frac{(\omega + \Omega_{\mu\alpha } - \bar{\lambda}_{\mu\alpha})^2}{ 4 \bar{\lambda}_{\mu\alpha} T} \right] + \exp\left(
\frac{\omega}{T}\right) \exp\left[ - \frac{(\omega - \Omega_{\mu\alpha } + \bar{\lambda}_{\mu\alpha})^2}{ 4 \bar{\lambda}_{\mu\alpha} T} \right]
\right\}. \label{GamTilde}
\end{eqnarray}
We note that $\tilde{\Gamma}_{\mu} = (1/2) \sum_{\alpha} \tilde{\gamma}_{\alpha \mu},$ and $\Omega_{\mu\nu} = \omega_{\mu\nu} - \lambda_{\mu} +
\lambda_{\nu}$ from Eq.~(\ref{OmegaMuNu}),(\ref{barlambda}).

Assuming that the environment fluctuations acting on each electron-binding site are independent and using Eq.~(\ref{LambdaMuNu}) for the
coefficients $\Lambda_j^{\mu\nu}$, we obtain
\begin{eqnarray}
\tilde{J}_{\mu\nu}(\omega) = \sum_l \left[ J_{lS}(\omega) |\langle \mu|S_l|\nu\rangle |^2 + J_{lM}(\omega) |\langle \mu|M_l|\nu\rangle |^2 \right] +
J_F(\omega) |\langle \mu|n_F|\nu\rangle |^2,
\end{eqnarray}
where
\begin{eqnarray}
 J_{lS}(\omega) = \sum_j \frac{m_j\omega_j^3}{2} \bar{x}^2_{jl} \delta(\omega - \omega_j), \nonumber\\
J_{lM}(\omega) = \sum_j \frac{m_j\omega_j^3}{2} \tilde{x}^2_{jl} \delta(\omega - \omega_j), \nonumber\\
J_F(\omega) = \sum_j \frac{m_j\omega_j^3}{2} x_{jF}^2 \delta(\omega - \omega_j).
\end{eqnarray}
The results obtained above are valid for an arbitrary frequency dependence of the spectral densities $J_{lS}(\omega), J_{lM}(\omega),
J_F(\omega) $. Hereafter we assume that these functions are described by the Lorentz-Drude formula characterized by a common inverse correlation
time, $\gamma_c = \tau_c^{-1}$, and by a corresponding reorganization energy $\lambda_{lS}, \lambda_{lM},$ or $\lambda_F$, e.g.
\begin{equation}
J_{lS}(\omega) = 2 \frac{\lambda_{lS}}{\pi} \frac{\omega\gamma_c}{\omega^2 + \gamma_c^2}.
\end{equation}

Quenching processes also contribute to the decay of the off-diagonal elements, $\langle \rho_{\mu\nu} \rangle$, with the following decoherence
rates: $ \Gamma_{\mu\nu}^{\rm quen} = \Gamma_{\mu}^{\rm quen} + \Gamma_{\nu}^{\rm quen},$ where
\begin{eqnarray}
\Gamma_{\mu}^{\rm quen} = \sum_{l\alpha} |\langle \mu |a_l^\dagger a_{l^*}| \alpha \rangle |^2 \chi''_l(\omega_{\mu\alpha} ) \left[
\coth\left(\frac{\omega_{\mu\alpha}}{2T}\right) + 1 \right]. \label{Gamquen}
\end{eqnarray}
Here we consider an Ohmic quenching heat-bath with the spectral density $\chi''_l(\omega ) = \alpha_l \omega$, which is determined by a set of
site-dependent dimensionless coupling constants $ \alpha_l \ll 1 $. The contribution of quenching to the relaxation of the diagonal elements of
the electron matrix, $\langle \rho_{\mu} \rangle$, is determined by the standard Redfield term
\begin{eqnarray}
\gamma_{\mu\nu}^{\rm quen} = \sum_l ( |\langle \mu |a_l^\dagger a_{l^*}| \nu \rangle |^2 + |\langle \nu |a_l^\dagger a_{l^*}| \mu \rangle |^2 )
\chi''_l(\omega_{\mu\nu} ) \left[ \coth\left(\frac{\omega_{\mu\nu}}{2T}\right) - 1 \right]. \label{gamquen}
\end{eqnarray}

As a result, we find that the time evolution of the off-diagonal elements of the electron matrix is determined by the expression
\begin{eqnarray}
\langle \rho_{\mu\nu}\rangle (t) = \exp\, (\, i\, \omega_{\mu\nu}\, t - \bar{\lambda}_{\mu\nu}\, T \,t^2\, ) \times \exp\,(\, -
\Gamma_{\mu\nu}\, t \,)\; \rho_{\mu\nu}(0), \label{rhoMuNuTime}
\end{eqnarray}
with the decoherence rates $\Gamma_{\mu\nu} = \Gamma_{\mu} + \Gamma_{\nu}$, where the coefficient $\Gamma_{\mu}$ contains contributions of the
off-diagonal fluctuations of the environment (\ref{GamTilde}) as well as quenching processes $\Gamma_{\mu}^{\rm quen}$ (\ref{Gamquen}):
$\Gamma_{\mu} = \tilde{\Gamma}_{\mu} + \Gamma_{\mu}^{\rm quen}$.  The evolution starts at the moment $t=0$ with the initial matrix
$\rho_{\mu\nu}(0)$. An effect of diagonal environment fluctuations is determined by the rate $\sqrt{\bar{\lambda}_{\mu\nu}\, T}$, where
$\bar{\lambda}_{\mu\nu}$ is the reorganization energy defined by Eq.~(\ref{barlambda}) and $T$ is the temperature of the environment.

\end{appendix}

\end{document}